\documentclass[apj]{emulateapj}
\usepackage{graphicx}
\usepackage{amsmath}
\usepackage{amssymb}
\usepackage{natbib}

\begin{document}
\title{Simulating nonlinear cosmological structure formation with massive
  neutrinos}

\author{Arka Banerjee \& Neal Dalal}
\affil{Department of Physics, University of Illinois at
  Urbana-Champaign, 1110 West Green Street,  Urbana, IL 61801-3080 USA}
\email{}

\begin{abstract}
We present a new method for simulating cosmologies that contain
massive particles with thermal free streaming motion, such as
massive neutrinos or warm/hot dark matter.  This method combines
particle and fluid descriptions of the thermal species to
eliminate the shot noise known to plague conventional N-body
simulations. We describe this method in detail, along with results
for a number of test cases to validate our method, and check its
range of applicability. Using this method, we demonstrate that 
massive neutrinos can produce a significant scale-dependence in the 
large-scale biasing of deep voids in the matter field.  We show that
this scale-dependence may be quantitatively understood using an
extremely simple spherical expansion model which reproduces the
behavior of the void bias for different neutrino parameters.
\end{abstract}

\section{Introduction}

One of the central tenets of the standard cosmological model is that
structure observed in the present universe formed via gravitational
evolution of initially linear density perturbations which arose
in the primordial universe \citep{Peebles1980,Dodelson2003book}.   
The intially linear perturbations responsible for
producing anisotropies in the cosmic microwave background
\citep{Planck2013} are expected to eventually develop into nonlinear
cosmological structures like halos and filaments at low redshift
\citep{Mo2010book}.  Predicting the nonlinear evolution of cosmic
structure has been a numerically challenging problem for many years. 
The method of choice for computing structure formation has been N-body
simulation.  The accuracy and efficiency of this method have been well
established for standard $\Lambda$CDM cosmologies
\citep{Heitmann2005,Heitmann2008}.  These simulations have been used
to study many aspects of structure formation, including the mass
functions \citep{Jenkins2001,Tinker2008,Warren2006} and profiles 
\citep{Navarro1997,Colberg2005,Zemp2009,Diemand2008,Diemand2011,Springel2008,
  Stadel2009} of halos and voids, and their large scale clustering
\citep[e.g.][]{Springel2005b,Gao2005,Tinker2008,Tinker2010}.

Although N-body simulations have been enormously successful 
in describing the evolution of $\Lambda$CDM cosmologies, 
they have not fared as well for studying certain other 
cosmologies, especially universes containing massive particles with large 
thermal velocities in the initial phase space distribution of the 
species. The large thermal velocities of these particles generate spurious 
structures in the density field due to discreteness effects inherent in N-body 
methods, as the particles stream large distances in random directions. At scales 
below this streaming scale, the density distribution of the simulation particles 
follows a Poisson shot noise distribution, instead of the correct, physical 
density. The power spectrum at large wavenumber would then be given by
$P(k) \sim \bar n ^{-1}$, 
where $\bar n$ is the mean density of particles in the simulation. For typical 
values of $\bar n$, the shot noise power spectrum can completely dominate the 
physical power spectrum (which is damped on these scales by the
streaming motions), leading to spurious structures forming everywhere
in the simulation volume. Therefore, improvements in numerical methods are  
essential to develop a reliable and consistent method for studying structure 
formation in such cosmologies. 

An example of a cosmology with fast moving massive particles is our
own universe, which is known to contain massive neutrinos.  Neutrinos
are among the most abundant particles in the universe, comparable to
photons in terms of their number density \citep{Dodelson2003book}.
Oscillation experiments have clearly established that at least two
neutrino states are massive, and these experiments have also placed
tight bounds on the mass differences between the three mass
eigenstates of neutrinos
\citep{SuperK98,SNO2001,K2K2003,Kamland2003,Dayabay2012}. Even though
the absolute masses are not yet known, the mass splittings imply
that at least one of the species is as heavy as $\sim0.06\,$eV. Given the
current cosmic neutrino background temperature $T_\nu\sim 1.68 \times
10^{-4}\,$eV \citep{Dodelson2003book}, this means that this species is
highly non-relativistic today, and can therefore gravitationally
cluster.  The effects of massive neutrinos on large-scale
structure are well-understood on large scales and at early times, when
fluctuations are in the linear regime
\citep{Lesgourgues2014}. However, at late times and smaller scales,
density fluctuations become non-linear,  rendering linear perturbation theory
calculations invalid.

Different groups have put forward different approaches to attacking this problem. 
\citet{Brandbyge2009} and \citet{Archidiacono2016} have proposed treating the neutrinos as a linear fluid on a 
grid coupled to the fully non-linear, N-body evolution of the cold
dark matter (CDM) particles. 
Similarly, \citet{Yacine2013} suggested solving the linearized Boltzmann 
equations for neutrinos coupled to the the N-body evolution of CDM particles. By 
their very nature, these quasi-linear methods break down when the overdensities 
in the neutrino fluid approach $\mathcal O (1)$. At late times, massive neutrinos 
can become cold enough to be captured into the deepest potential wells 
in the simulations - the largest halos.  The overdensities of 
neutrinos in these halos can be significantly larger than $\mathcal O(1)$, 
limiting the validity of linearized methods at late times.

Another method which has been proposed is to treat the neutrinos as a different 
species of particles with different mass than CDM particles in a
normal N-body simulation at all redshifts \citep{Villaescusa2014}. 
A related method is to use the linear 
treatment for neutrinos at early redshifts, but then to switch to N-body 
treatment once perturbations in the neutrino fluid become non-linear 
\citep{Brandbyge2010a}.  While traditional N-body methods suffer from the shot 
noise issue mentioned above, these authors argue that since the neutrinos 
constitute a small part of the energy budget ($\Omega_\nu \ll 1$), their effects 
are always subdominant to CDM, and the shot noise effect is not strong enough to 
significantly alter structure formation.  On the other hand, since
shot noise can be significant compared to the real physical clustering
of neutrinos in the simulation, this means that calculations of
neutrino effects on the power spectrum can suffer from large {\em
  fractional} errors, even if the absolute errors on the power
spectrum are small thanks to the small mass fraction in neutrinos.
Given that the point of performing such simulations is the precise
calculation of neutrino effects, the fractional error on neutrino
effects may be a more relevant metric than the absolute error on the
total power spectrum.  A number of interesting results 
have been found using these methods \citep{Villaescusa2011, 
Villaescusa2013,Brandbyge2010b,Castorina2014,Costanzi2013,Inman2015,
Castorina2015,Carbone2016}. However, with future 
cosmological surveys expected to reduce error bars on multiple observables 
significantly, an improvement in the accuracy of neutrino simulations
may now be warranted.

Similarly, besides neutrinos, dark matter particles can also have significant random 
thermal velocities, depending on the DM temperature at freezeout.  It has been 
suggested that Warm Dark Matter \citep{Bode2001} can alleviate certain 
small-scale problems present for CDM universes, like the core-cusp problem of 
halo profiles \citep{deBlok2008,deBlok2010} and the missing satellite problem 
\citep{Moore1999,Klypin1999,Diemand2008,Springel2008}.  If DM particles have a 
finite thermal velocity dispersion, then those particles will randomly stream a 
finite distance, and this random streaming acts to suppress structure on scales 
below the free-streaming length.  The streaming length of WDM particles is much 
smaller than the streaming length of neutrinos 
\citep{Bode2001,Villaescusa2011b}, and if the simulations are initialized in a way 
that the particles have random thermal velocities, the scales affected by shot 
noise would be smaller than in neutrino simulations. We refer to these sort of 
simulations as ``hot start'' simulations. However, since the WDM is the dominant 
component in these simulations, any amount of shot noise coming from the thermal 
velocities is sufficient to seed the formation of spurious structures. ``Hot 
start'' N-body simulations are therefore not an accurate method for studying 
such WDM cosmologies.

To get around this problem, the ``cold start'' method has typically been used in 
the literature \citep{AvilaReese2001,Wang2007}. In this method, the random 
thermal velocities of particles are {\em not} included in the simulation initial 
conditions.  To account for the damping of the power spectrum, this method 
initializes using the linear power spectrum for the WDM species at redshift 
$z=0$, scaled back to the starting redshift of the simulation using CDM growth 
factors.  This method therefore necessarily does not capture the spatial 
dependence and time dependence of the growth of structure, but it {\em does} 
eliminate artifacts arising from random thermal velocities.  Nonetheless, the 
cold start method suffers from its own artifacts, like the ``beads on a string'' 
phenomenon \citep{Wang2007}.  \citet{Angulo2013} have proposed an 
interesting method to eliminate these ``beads on a string'' artifacts and other 
spurious structures from WDM simulations. Their method employs a tetrahedral 
tessellation of the 6-dimensional phase space of  simulation particles to follow 
the evolution of the WDM densities. However, it is not yet clear if this method 
is accurate inside collapsed and virialized regions such as halos or subhalos, 
whose abundance will likely provide stringent constraints on WDM models in 
upcoming years \citep[e.g.][]{Hezaveh2013,Hezaveh2016}.
Alternatively, \citet{Hobbs2016} have proposed that adaptive softening
of the gravitational force can help to suppress spurious structure
found in cold start simulations of WDM cosmologies.  

In this paper, we present a novel method to simulate cosmologies with hot 
particles which is valid at all redshifts, in both linear and non-linear 
regimes. This method makes use of both particle techniques from N-body 
simulations as well as hydrodynamic techniques from fluid simulations. 
This paper is arranged as follows. In \S \ref{eom}, we derive the relevant 
equations of motion for hot species in an expanding universe.  We describe the 
implementation of our method at early and late redshifts in \S 
\ref{closehierarchy}. In \S \ref{Methods} we discuss the time integration 
techniques for the particles in our simulations, as well the hydrodynamic scheme 
we implement. In \S \ref{codetest}, we discuss a number of tests, in both the 
linear as well as the non-linear regime to validate our code. We use this method 
to show a novel effect in void biasing with neutrinos in \S \ref{nusim}. 
Finally, we list our main conclusions and directions for future work in \S 
\ref{concl}.

\section{Equations of motion}
\label{eom}

In this section, we review the equations of motion governing hot species in an 
expanding universe. We take moments of the Boltzmann equation to derive 
effective fluid equations that allow us to evolve the hot species using 
hydrodynamic methods.

\subsection{Collisionless Boltzmann equation}
\label{CBE}
For the applications we are interested in, we consider the WDM and neutrinos to 
be essentially collisionless, and our starting point will be the collisionless 
Boltzmann equation 
\begin{equation}
\frac{df}{dt} = \frac{\partial f}{\partial t} + \frac{\partial f}{\partial 
x^i}\frac{d x^i}{dt}
			+ \frac{\partial f}{\partial p^i}\frac{dp^i}{dt} = 0
\label{boltzmann}
\end{equation}
where $f(\mathbf x, \mathbf p)$ is the phase space distribution function of the 
dynamical species in the simulation volume.  Working in Newtonian gauge with 
the Newtonian potentials $\Phi$ and $\Psi$, which are typically of order 
$10^{-5}$  in units where $c=1$, we have up to first order in the potentials 
\citep{Dodelson2003book}
\begin{equation}
\frac{d x^i}{dt} = \frac{p^i}{a E} \left(1 - \Phi + \Psi \right) \,.
\label{velocity}
\end{equation}
In equation \eqref{velocity}, $E = \sqrt{p^ip^i+m^2}$ and $a$ is the scale 
factor. Similarly, 
\begin{equation}
\frac{d p^i}{dt}  = - p^i \frac{\partial \Phi}{\partial t} - p^i \frac{\dot 
a}{a} 
	- \frac{E}{a} \frac{\partial \Psi}{\partial x^i}
\label{acceleration}
\end{equation}
to first order in the potentials. Here $\dot a$ means a derivative with respect to time,
and not a conformal time derivative. Substituting \eqref{velocity} and 
\eqref{acceleration} into \eqref{boltzmann} we have
\begin{eqnarray}
\label{boltzmann2}
\frac{\partial f}{\partial t} &+& \frac{\partial f}{\partial 
x^i}\left[\frac{p^i}{a E} \left(1 - \Phi + \Psi \right)\right]
			\nonumber 
			\\ &+& \frac{\partial f}{\partial p^i}\left[- p^i 
\frac{\partial \Phi}{\partial t} - p^i \frac{\dot a}{a} 
	- \frac{E}{a} \frac{\partial \Psi}{\partial x^i}\right]= 0 \,.
\end{eqnarray}

\subsection{Poisson equation}
The metric potentials $\Phi$ and $\Psi$ are related to the matter content via 
the Poisson equation.  
Since we are interested in epochs during matter domination ($z\lesssim 300$) 
when the overall anisotropic stress is small, we can assume $\Psi = - \Phi$. In 
Newtonian gauge, we have
\begin{equation}
\label{Poisson}
\nabla^2 \Psi = 4\pi G a^2 \left[\bar \rho \delta - 3 \frac{\dot a}{a}(1+ w) 
\partial_i v^i\right]
\end{equation}
where $v_i$ is the local peculiar velocity, $\delta=\delta\rho/\bar\rho$ is the 
local overdensity, and $w=\bar P/\bar\rho$ is the background equation of state.  
All of these fluid quantities will be defined formally in terms of moments of 
the distribution function below.
For simulation boxes where the volume is much smaller than the Hubble volume, 
the second term in the brackets on the right hand side of eqn.\ \eqref{Poisson} can be 
neglected.

\subsection{Obtaining the Boltzmann moment equations}
\label{BME}
We can use the Boltzmann equation \eqref{boltzmann} to derive fluid equations 
for collisionless particles by integrating over various moments.  First, we 
derive the continuity equation in the usual way, by multiplying 
\eqref{boltzmann2} by $E$ and integrating over momentum. We define the 
real-space density in terms of the phase space density in the following manner, 
\begin{equation}
\rho(\mathbf x) = \int E f(\mathbf x, \mathbf p) d^3 \mathbf p
\label{density}
\end{equation}
Since we are interested in the relative density contrast $\delta = (\rho - \bar 
\rho)/\bar \rho$, we can cast the continuity equation into an equation for the 
time evolution of $\delta$:
\begin{eqnarray}
\label{continuity}
\dot \delta = - \frac{1}{a}\frac{\partial \left[(1+2\Psi )\Pi^i\right]}{\partial 
x^i} 
&-&3\dot \Phi (1+\delta)(1+{W}) \nonumber \\
&-& 3 \frac{\dot a}{a}(1+\delta)\left({W} - w\right) \,.
\end{eqnarray}
where $w = \bar P /\bar \rho$ is the background equation of state and we have 
defined 
\begin{equation}
\label{momentum}
\Pi^i (\mathbf x)= \frac{\displaystyle \int p^i f(\mathbf x, \mathbf p) d^3 
\mathbf p}{\bar \rho}
\end{equation}
and 
\begin{equation}
W(\mathbf x)= \frac{\displaystyle \int \frac{p^ip^i}{3 E} f(\mathbf x, \mathbf 
p) d^3 \mathbf p}{\rho(\mathbf x)} .
\end{equation}
If the species is non-relativistic, the bulk velocity can be simply defined as
\begin{equation}
\label{bulkvelocity}
v^i(\mathbf x) = \frac{\Pi^i(\mathbf x)}{1+\delta(\mathbf x)}
\end{equation}
In eqn.\ \eqref{continuity}, the last two terms on the right hand side are 
typically small compared to the first term. Since we are interested in matter 
domination regimes, $\dot \Phi $ is small compared to the spatial derivatives of 
$\Phi$, and we can neglect the second term on the RHS without a loss of 
accuracy. If the particles are relativistic, the last term is small or zero 
because the local sound speed and the background sound speeds are the same, and 
$(W- w)$ vanishes. When particles are non-relativistic, $W$ and $w$ are 
individually small ($\sim 10^{-6}$) and even in non-linear regimes the last term 
remains much smaller than the first term.

Next we multiply eqn.\ \eqref{boltzmann2} by $p^i$ and integrate to get the 
Euler equation for the fluid
\begin{eqnarray}
\label{euler}
\dot{ \Pi^i} = - \left(1 - 3 w\right)\frac{\dot a}{a} \Pi^i &-& 
\frac{1}{a}\frac{\partial \left[(1+\delta)W^{ij}\right]}{\partial x^j}\nonumber 
\\
 &-& \frac{1+\delta}{a} \frac{\partial \Psi}{\partial x^i} 
\end{eqnarray}
with 
\begin{equation}
\label{dispersion}
 W^{ij}(\mathbf x) = \frac{\displaystyle \int \frac{p^ip^j}{E}f(\mathbf x, 
\mathbf p)d^3\mathbf p}{\rho(\mathbf x)}
\end{equation}
Note that, in principle, we can continue this procedure of generating equations 
using higher and higher moments of the Boltzmann equation. Each equation will be 
coupled to the next - this is apparent by looking at the structure of the two 
equations we have derived, \eqref{continuity} and \eqref{euler}. In the next 
section, we comment on how we close this infinite hierarchy of equations.
 
\subsection{SPH equations}

As we will show in \S\ref{codetest}, there are some situations in which it is 
advantageous to use a Lagrangian description of the fluid rather than an 
Eulerian description. We use a Smoothed Particle Hydrodynamics (SPH) approach for these 
problems, following the procedures in \citet{Monaghan1992}. For non-relativistic 
collisionless particles, the equations of motion can be written as 
\begin{equation}
\label{sph_x}
\frac{d x^i}{dt} = \frac{v^i}{a}\left(1-\Phi + \Psi\right)
\end{equation}
and
\begin{eqnarray}
\label{sph_v}
\frac{d v^i}{dt} = &-&v^i \frac{\partial \Phi}{\partial t} - v^i \frac{\dot 
a}{a} - \frac{1}{a}\frac{\partial \Psi}{\partial x^i}\nonumber \\ &-& 
\frac{1}{\left(1+\delta\right)a}\frac{\partial\left(\left(1+\delta\right) 
W^{ij}\right)}{\partial x^j} .
\end{eqnarray}

\section{Closing the hierarchy}
\label{closehierarchy}
To study the evolution of the collisionless fluid with the above equations, we 
need some way to close the Boltzmann hierarchy. For collisional fluids, one can 
use an equation of state to relate the energy density and the pressure to close 
the system of equations. However, in the collisionless cases that we will be 
interested in (neutrinos and WDM) there is no simple equation of 
state, and so an alternative closure method is required.

To motivate our method to close the Boltzmann hierarchy, it will be useful to 
consider the approach used in particle-mesh (PM) N-body simulations 
\citep{Hockney1988}.  In N-body simulations, Lagrangian particles are evolved 
under the influence of their collective gravitational field.  Those particles 
represent a Monte Carlo sampling of the distribution function, and in PM 
simulations, those particles are used to estimate the density field 
$\delta(\mathbf x)$ that enters Eqn.\ \eqref{Poisson}.  In effect, PM 
simulations use particles to close the Boltzmann hierarchy at its zeroth moment. 
 In principle, however, we could use those particles to estimate other 
quantities that enter the fluid equations.  For example, we could use particles 
to estimate a bulk fluid velocity, and then use that estimated velocity to 
evolve the density field using the continuity equation \eqref{continuity}.  
Alternatively, we could use particles to estimate a stress tensor entering the 
Euler equation \eqref{euler} that would truncate the hierarchy at its 2nd 
moment.  Indeed we can estimate an arbitrary moment of the distribution function 
from particles, and truncate the Boltzmann hierarchy accordingly.

Therefore, the method we use is the following.  We represent the collisionless 
fluid (e.g.\ neutrinos or WDM) simultaneously using fluid quantities on a grid 
and using test particles as well.  We evolve the grid fluid using fluid 
equations like \eqref{continuity} and \eqref{euler}, and we truncate the 
Boltzmann hierarchy of fluid equations using moments of the distribution 
function estimated from the test particles, e.g.\ Eqn.\ \eqref{momentum} or 
\eqref{dispersion}.  The test particles evolve under the gravitational field 
estimated from the fluid, i.e.\ Eqns.\ \eqref{velocity} and 
\eqref{acceleration}.  Compared to traditional PM simulations, our approach 
involves solving more equations than the Poisson equation (i.e.\ fluid 
equations), and involves estimating higher moments of the particle distribution 
function (i.e.\ 3 components of the bulk velocity, or 6 components of the stress 
tensor, rather than a single scalar density field).  This would appear to be 
considerably more expensive than traditional PM codes, but as we argue below, 
the benefits of using this approach in certain situations can outweigh the added 
costs.

\subsection{Early evolution}
\label{early_ev}
We initialize our simulations using perturbation theory.  We use Eulerian PT to 
initialize fluid quantities on the grid, and Lagrangian PT to initialize the 
test particles.  In addition to the LPT velocities, the test particles are also 
given random thermal velocities drawn from the distribution function of the 
species we are interested in. 
 
At early times, the particles can have large thermal velocities. These random 
thermal velocities can produce shot noise in any quantity we try to estimate 
from the particles, in the same way that the shot noise in particle positions 
generates noise in the density field computed from the particles, as discussed 
in the introduction.  One difference between the shot noise in particle 
velocities, compared to the noise in particle positions, is that the velocity 
noise diminishes over time as the universe expands.  This means that shot noise 
in fluid quantities like the bulk velocity or velocity dispersion becomes small 
at low redshift, in contrast to the shot noise in the estimated density field.  
This illustrates one reason why it can be advantageous to estimate quantities 
other than the density from the particles.

Nevertheless, at early times the shot noise in velocities is large.  In 
principle, this can be suppressed by increasing the number of particles in the 
simulation, but in many cases of interest, the required number of particles is 
orders of magnitude too large to be feasible.   Therefore, using more particles 
in the simulation is not a practical solution in most situations of interest.  

Another way to effectively increase the number of particles used in estimating 
fluid quantities is to spatially smooth those quantities.  Spatially smoothing 
the fluid quantities is equivalent to estimating those quantities at a point 
using a larger volume, and hence more particles.  The obvious reason why 
simulations normally do not spatially smooth over large volumes to suppress shot 
noise is that smoothing erases any small-scale structure in the estimated 
quantities.  For species like cold dark matter, there is structure on all 
spatial scales, and so spatially smoothing would incorrectly eliminate real 
physical structure in the DM distribution.  However, for the hot species of 
interest to us, the high temperature implies that there is a Jeans scale $k_J 
\sim aH/c_s$ below which small-scale structure is actively suppressed.  
Arguably, therefore, spatially smoothing is not necessarily invalid as long as 
the smoothing length is always safely below the Jeans scale.

On the other hand, the power spectrum does not vanish at $k>k_J$.  Since we 
would like to accurately evolve the power spectrum on all scales in the 
simulation, including scales below the Jeans scale, this restricts what fluid 
quantities we can spatially smooth.  For example, we cannot use the smoothing 
technique to estimate the velocity field, because this will extinguish the growth of 
structure on all scales below the smoothing scale. This becomes apparent if one 
looks at the linear continuity equation in Fourier space:
\begin{equation}
\label{linear_continuity}
\dot \delta_k = - \frac{i \mathbf k. \mathbf v_k}{a}
\end{equation} 
Smoothing sets $\mathbf v_k$ to $0$ for $k>k_{\rm smoothing}$, which gives the 
unphysical result $\dot \delta_k = 0$.

However, from our tests, we find that we {\em can} spatially smooth the velocity dispersion $W^{ij}$ at early times without sacrificing accuracy.  To see this, note that when the 
thermal velocities are large enough to require spatial smoothing to suppress 
shot noise, then the Jeans scale is large, and so structure in the hot species 
remains linear.  Under the fluid approximation, fluctuations in the velocity dispersion are second order, 
however, since they involve two perturbed velocities.  Therefore, spatially 
smoothing the velocity dispersion only drops second order fluctuations in the fluid approximation, and does 
not affect the linear evolution of the velocity field in the Euler equation 
\eqref{euler}.  This breaks down when structure in the hot species becomes 
nonlinear, but in order for structure to become nonlinear, the thermal 
velocities must be small, eliminating the need for spatial smoothing to suppress 
shot noise.  Therefore, at all redshifts, we can estimate fluid quantities 
without significant thermal shot noise.  

In order for this method to work, we must pick a sensible smoothing length.  If 
the smoothing length is too small, shot noise will corrupt the evolution of 
fluid quantities, whereas if the smoothing length is too large (exceeding the 
Jeans scale), then spatial smoothing artificially removes real physical 
structure in the simulation.  We set the smoothing length using the following 
argument.  The quantity we are estimating from the particles is the velocity 
dispersion, and we require that the error in our estimate of the dispersion to 
be small compared to velocities generated by gravity.  
We estimate the dispersion at every point on the grid by evaluating 
\begin{equation}
\label{disp_definition}
 W^{ij} (\mathbf x)= \frac{\sum_{\mathbf x} \displaystyle 
\frac{p^ip^j}{E}}{\sum_{\mathbf x} E} - \frac{\sum_{\mathbf x} 
p^i}{\sum_{\mathbf x} E } \frac{\sum_{\mathbf x} p^j}{\sum_{\mathbf x} E } 
\end{equation}
where $\sum_{\mathbf x}$ stands for the sum over all particles at position 
$\mathbf x$. Note that for non-relativistic particles, $E$ for every particle is 
approximately the mass, and so the denominators in the above expression count 
the total mass of particles at a given point. We use a cloud in cell (CIC) 
scheme to evaluate the different sums. 
The average thermal velocity dispersion is given by the equation of state $w$. 
Therefore, if $N$ particles have been used to estimate the velocity dispersion, 
the error in the estimate will be $\Delta = w/\sqrt N$. The error is going to be 
small if $\Delta$ is small compared to the average velocity dispersion sourced 
by gravity $\langle v_{esc}^2 \rangle$. The latter can be estimated by 
evaluating the average of the magnitude of the gravitational potential in the 
box, $|\Phi_{rms}|$. If we set $\Delta = \epsilon |\Phi_{rms}|$ for some error 
tolerance $\epsilon \ll 1$, then the number of particles that are needed to 
achieve the error tolerance is
\begin{equation}
N = \frac{w^2}{\epsilon^2 |\Phi_{rms}|^2}.
\end{equation}
When structure is linear, we know that $n$, the average number of particles per 
pixel, is a good estimate of the actual number of particles per pixel (modulo 
shot noise).  This means that we have 
\begin{equation}
N = n V_s \approx n L_s^3
\end{equation}
where $V_s \propto L_s^3$ is the smoothing volume required. This gives us an 
expression for the smoothing length 
\begin{equation}
\label{smoothing_length}
L_s =  \left(\frac{w^2}{\epsilon^2 |\Phi_{rms}|^2 n} \right)^{1/3}.
\end{equation}

We can adjust $\epsilon$ and $n$ to ensure that our smoothing length never 
exceeds the local Jeans scale calculated in linear theory. We also need to 
adjust our parameters so that we are no longer smoothing when non-linear 
structures like halos start forming in the simulation volume. At early times, 
the particles are hot (large $w$) and the smoothing length is large. As time 
progresses, the particles become colder, which means $w$ decreases. At the same 
time, structure starts forming in the box and $|\Phi_{rms}|$ increases with 
time. These effects together mean that the smoothing length is a rapidly 
decreasing function of time.

\subsection{Late evolution}

As the simulation proceeds, the smoothing length reduces below a pixel at some 
redshift. The random motions of particles at redshifts after this time do not 
produce levels of shot noise which will affect the evolution of the power 
spectrum.  Once the shot noise becomes negligible, we do not need to spatially 
smooth the fluid quantities.  At subsequent times, our fluid approach may appear 
unnecessary, given the computational costs associated with our method compared 
to traditional N-body methods.  Unfortunately, we cannot switch from fluid 
evolution to N-body even when the smoothing length is less than 1 pixel.  The 
reason is that the particle density field has shot noise in it even at late 
time, arising from the motion of the particles at higher redshifts.  Using the 
density field for further evolution would mean that structure induced by this 
shot noise would grow gravitationally, and would start forming spurious halos at 
lower redshifts. This means that even at later times, the density field of the 
particles should not be used directly to source gravity.

However, we can safely switch from estimating the velocity dispersion and 
evolving the fluid density and bulk velocity fields, to estimating the bulk 
velocity field and only evolving the density field using the continuity equation 
\eqref{continuity}.  This is safe at late times because we are not spatially 
smoothing the estimated velocity field.  Switching from estimating dispersion to 
estimating velocity produces a considerable speed-up in the simulation, since 
fewer quantities are being estimated (3 velocities vs. 6 dispersions) and fewer 
fluid equations need to be evolved (only continuity, not Euler).  To estimate 
bulk velocities on the grid from the positions and velocities of the particles, 
we once again use a CIC interpolation scheme,  
\begin{equation}
u^i(\mathbf x)= \frac{\sum_{\mathbf x} p^i}{\sum_{\mathbf x} E}.
\end{equation}
As noted above, switching to velocity estimation speeds up the code 
significantly. We have verified that we obtain consistent simulation results 
using the faster velocity estimation and the slower dispersion estimation at 
late times.  Therefore, in all of our Eulerian simulations below, we will switch 
to velocity estimation once the smoothing length shrinks to below one pixel.

\subsection{Smoothing in SPH simulations}
\label{sphmethod}

We will show below in \S\ref{codetest} that in certain cases, it can
be advantageous to use a Lagrangian formulation of the fluid equations
rather than an Eulerian description.  The Lagrangian description we
will use in those cases is smoothed particle hydrodynamics (SPH).  
In those simulations, we use the following technique. We 
use two sets of particles - the first set evolves following the SPH equations of 
motion Eqs.\ \eqref{sph_x}, \eqref{sph_v}, and a second set of test particles 
which are evolved using Eqs.\ \eqref{velocity}, \eqref{acceleration}. Like in the 
Eulerian case, the test particles are given thermal velocities drawn from their 
Fermi-Dirac distributions. Once again, these particles are not used in any of 
the density estimates, but only used to measure the velocity dispersion on a 
grid, as defined in Eqn.\ \eqref{disp_definition}. Such an estimate of the velocity 
dispersion will be plagued by shot noise at early times, and needs to be 
smoothed, and this is done following the same prescription for the smoothing 
length discussed in \S\ref{early_ev}. Given the smoothed velocity 
dispersions on a grid, we interpolate from the grid to the positions of 
individual SPH particles to assign velocity dispersions to them. Once the 
thermal dispersions have been assigned, we use standard cubic spline 
interpolations from \citet{Monaghan1992} to measure the density and the velcoity 
dispersion gradient required in Eq.\ \eqref{sph_v}. For example, the density the 
position of the $a$-th particle is given by \citep{Monaghan1992}
\begin{equation}
\label{sphdensity}
\rho(\mathbf r_a) = \Sigma_b m_b W(\mathbf r_a - \mathbf r_b, h)
\end{equation}
where the sum runs over all other particles, $W$ is the interpolation kernel, 
and $h$ is the spline smoothing length, which we choose to be one grid cell. 
Notice that this spline smoothing length is different from the smoothing length 
we defined in Eq.\ \eqref{smoothing_length}.

Once the smoothing length from Eq.\ \eqref{smoothing_length} falls below our 
force resolution on the grid, we start treating the SPH particles as standard 
N-body particles for the rest of the evolution, as the effects of their thermal 
velocities beyond that point is below the resolution of the simulation. In 
practice, this means that we use Eqs.\ \eqref{velocity}, \eqref{acceleration} for 
time evolution.

Our technique does not make full use of the capabilities of SPH, as we use an 
intermediate grid to find and smooth the velocity dispersions from the test 
particles. This automatically limits the resolution of the SPH technique, but 
since our objective was only to suppress the shot noise from thermal velocities and 
given that our gravitational force resolution is also limited by the same grid, 
the above method is sufficient for the purposes of our simulations.

\section{Integration techniques}
\label{Methods}
\subsection{Particle time integration}
\label{particlemethod}
For the test particles we use in the simulations, as well as for CDM particles 
in neutrino simulations, we use the standard Kick-Drift-Kick leapfrog time 
integration \citep{Springel2005}:
\begin{eqnarray}
\mathbf v ^{n+\frac 12} &=& \mathbf v^ n + \frac{\Delta t}{2} \mathbf f^n \\
\mathbf x ^{n+1} &=& \mathbf x^n + \Delta t\, \mathbf v^{n + \frac 12} \\
\mathbf v ^{n+1} &=& \mathbf v^{n+\frac 12} + \frac{\Delta t}{2} \mathbf f^{n+1}
\end{eqnarray}
where $\mathbf x^i$, $\mathbf v^i$ are the particle positions and velocities at 
time step $i$, and $\mathbf f^i$ are the forces at those timesteps. For the 
Kick-Drift-Kick method, the Poisson equation is solved after the particle or 
Drift update - with the updated positions of the particles. This method is 
formally second order accurate, apart from being symplectic in nature 
\citep{Springel2005}.

The Drift-Kick-Drift method \citep{Springel2005}
\begin{eqnarray}
\mathbf x ^{n+\frac 12} &=& \mathbf x^n + \frac{\Delta t}{2} \mathbf v^{n} \\
\mathbf v ^{n+1} &=& \mathbf v^ n + {\Delta t} \, \mathbf f^{n+\frac 12} \\
\mathbf x ^{n+1} &=& \mathbf x^{n+\frac 12} + \frac{\Delta t}{2} \mathbf v^{n + 
1} 
\end{eqnarray}
is also second order accurate and symplectic, but we use the Kick-Drift-Kick 
method because the latter uses the forces (and hence the potential) at full time 
steps, whereas the former uses the potential at half time steps. The test 
particles are going to be coupled to a fluid, whose own time integration scheme 
gives the potential at full time steps, and so using Kick-Drift-Kick is 
essential for the fluid and the particles to remain coupled to each other.

\subsection{Hydrodynamics}
\label{hydromethod}
As we saw earlier, we will be solving the continuity and Euler equations at 
early times when the smoothing length defined by Eq.\ \eqref{smoothing_length} is 
larger than a grid cell, and most structure in the box is linear. At late times, 
when the smoothing length falls below a grid cell and highly non-linear 
structures start forming in the box, we will be solving only the continuity 
equation. Both equations are  hyperbolic partial differential equations with 
source terms.

We use an operator splitting method to split any source term present in the 
equations from the hyperbolic advection part. We also use directional splitting 
\citep{Toro2009} so that the advection in 3 dimensions is reduced to 3 
1-dimensional advections. To solve the 1-dimensional advection problem on the 
grid, we employ a finite-volume scheme which is piecewise linear, and hence 
second order accurate in space. 

The advection equation
\begin{equation}
\frac{\partial q}{\partial t} = -\frac{\partial (q \, u)}{\partial x}
\end{equation}
is discretized so that the update equation for $q^n_i$, the value of $q$ at cell 
center $i$ at timestep $n$ can be written as 
\begin{equation}
\frac{q^{n+1}_i-q^n_i}{\Delta t} = - \frac{f^{n+\frac 12}_{i+\frac 
12}-f^{n+\frac 12}_{i-\frac 12}}{\Delta x}
\end{equation}
where $f^{n+\frac 12}_{i\pm\frac 12}$ are the fluxes at the cell edges at 
timestep $n+\frac 12$ constructed from the data at timestep $n$. The fluxes are 
constructed in the following manner:
\begin{eqnarray}
f^{n+\frac 12}_{i+\frac 12} &=& 0.5 \, u^n_{i+\frac 
12}\left(\left(1+\theta_{i+\frac 12}\right)q^n_{i}
+\left(1-\theta_{i+\frac 12}\right)q^n_{i+1})\right)\nonumber \\
&+& 0.5 \, \left| u^n_{i+\frac 12}\right | \left(1-|c_{i+\frac 12}|\right) 
\phi(r^n_{i+\frac 12}) (q^n_{i+1}-q^n_i)
\end{eqnarray}
\begin{eqnarray}
f^{n+\frac 12}_{i-\frac 12} &=& 0.5 \, u^n_{i-\frac 
12}\left(\left(1+\theta_{i-\frac 12}\right)q^n_{i-1}
+\left(1-\theta_{i-\frac 12}\right)q^n_{i})\right)\nonumber \\
& +& 0.5 \, \left| u^n_{i-\frac 12}\right | \left(1-|c_{i-\frac 12}|\right) 
\phi(r^n_{i-\frac 12}) (q^n_{i}-q^n_{i-1})
\end{eqnarray}
where $c_i = u_i \Delta t/ \Delta x$. We define $\theta_{i\pm\frac 12} =1$ for 
$u_{i\pm \frac 12}>0$ and $\theta_{i\pm\frac 12} =-1$ for $u_{i\pm \frac 12}<0$. 
We further define
\begin{equation}
r^n_{i-\frac 12} = \begin{cases} \frac{q^n_{i-1}-q^n_{i-2}}{q^n_{i}-q^n_{i-1}} & 
\mbox {if} \,u_{i-\frac 12}>0 \\
\frac{q^n_{i+1}-q^n_{i}}{q^n_{i}-q^n_{i-1}} & \mbox {if} \,u_{i- \frac 12}<0
\end{cases}
\end{equation}
and similarly for $r^n_{i+\frac 12}$. $\phi(r)$ is the flux limiter function 
which is required so that the method is Total Variation Diminishing 
\citep{Harten1983}, by converting to a first order method near extrema in the 
profile of $q$. Even though there are no real shocks in the collisionless fluids 
that we will be dealing with, once non-linear structures start forming in the 
box, there are sharp density gradients, which lead to spurious oscillations if 
the hydro scheme we use is not TVD in nature. Though the TVD property is 
essential for the stability of the code, it also means that there will be 
artificial diffusion near the extrema - sharper the change in gradient,larger 
the diffusion. In our code tests and cosmological simulations we use the 
Superbee flux limiter \citep{Roe1986}, defined as
\begin{equation}
\label{superbee}
\phi(r)= \begin{cases}
0 & \mbox{if} \, r<0 \\
\min(2r,1) & \mbox{if} \, 0<r<1 \\
\min(2,r) & \mbox{if} \, 1<r
\end{cases}
\end{equation}
This turns out to be the least diffusive flux limiter among the ones we tested.

While the piecewise linear method is formally correct to second order in time in 
smooth regions, it switches to first order time accuracy near saddle points and 
extrema. In cosmological simulations, especially the ones involving WDM, saddle 
points appear throughout the box as structures form and move under the influence 
of gravity. This means that, typically, the solution will only be first order 
accurate in large parts of the box. To make the scheme at least second order 
accurate in time throughout the box, we use a second order explicit Runge-Kutta 
time integration scheme. We use data at timestep $n$ for the predictor step to 
get the predicted data at time $n+1$, and then use this information in the 
corrector step to get the final solution at time $n+1$. As we mentioned in the 
previous subsection, we calculate fluid quantities like the density at every 
full timestep, and therefore, to keep the particles and the fluid coupled, we 
use the Kick-Drift-Kick method which requires forces and potentials at every 
full time step, rather than the Drift-Kick-Drift method.

\subsection{Gravity}

We use Fast Fourier Transforms (FFT) to determine the potential on the grid  
from the densities of the neutrino fluid and WDM or CDM particles,
just as in PM simulations \citep{Hockney1988}. For CDM, we 
use cloud-in-cell (CIC) interpolation to obtain the grid density from the 
positions of the particles. In Fourier space, the Poisson equation is given by 
\begin{equation}
\label{fourierpoisson}
\tilde{\phi}(\mathbf k) = G(\mathbf k) \tilde{\delta}(\mathbf k)
\end{equation}
where $G(\mathbf k)$ is the Green function for the Poisson equation. On the 
grid, the discrete version of $G(\mathbf k)$ becomes
\begin{eqnarray}
G(k_x,k_y,k_z) = &-&C\bigg[\sin^2\left(\frac {k_x} 2\right)\nonumber \\
&+&\sin^2\left(\frac {k_y} 2\right)+\sin^2\left(\frac {k_z} 2\right)\bigg]^{-1}
\end{eqnarray}
where $C$ is a constant independent of scale and $\{k_x,k_y,k_z\}$ are the 
wavenumbers on the cubic grid.
Once we solve the Poisson equation using the FFT method, we calculate forces on 
the grid using a four point stencil
\begin{eqnarray}
f_x (i,j,k)&=& - \frac{\partial \phi_{(i,j,k)}}{\partial x} \nonumber \\
&=& - \frac{1}{12 \Delta  x}\bigg[8 \phi_{(i+1,j,k)} - 8 \phi_{(i-1,j,k)} 
\nonumber \\
&&- \phi_{(i+2,j,k)} + \phi_{(i-2,j,k)}\bigg]
\end{eqnarray} 
where $i$, $j$ and $k$ label the coordinates of the grid points. These forces 
are then used to update the fluid when we solve the Euler equation. For updating 
the velocities of the particles, we again use a CIC interpolation to interpolate 
the forces from the grid to the positions of individual particles. 

\section{Code tests}
\label{codetest}
In this section, we present various tests of the new method described
above. The first two tests are  
designed to check the accuracy of this method for the dynamics of virialized 
objects.  The third test is devised to check accuracy at early times, when 
traditional N-body simulations can produce large errors. We then compare the results
of this method to N-body results for a $\Lambda$CDM universe, where the we know the latter
yields accurate results. In our final test, we run simulations of Warm
Dark Matter (WDM) 
cosmologies to check whether this method is able to eliminate the spurious halos that are known 
to plague both ``hot start'' and ``cold start'' N-body simulations of WDM cosmologies.

\subsection{Plummer sphere advection}
\label{plummerad}
The Plummer sphere has an isotropic mass distribution with a radial 
density profile given by 
\begin{equation}
\label{plummer_density}
\rho(r) = \left(\frac{3M}{4\pi b^3}\right)\left(1+ \frac{r^2}{b^2}\right)^{-5/2}
\end{equation}
where $b$ is the Plummer radius, which sets the size of the virialized region. 
The potential for the Plummer sphere is
\begin{equation}
\label{plummer_potential}
\Phi(r) = - \frac{GM}{\left(r^2+b^2\right)^{1/2}}
\end{equation}
For this density profile, the phase space distribution function
$f(\mathbf r, \mathbf v)$ is \citep{binney2011galactic}  
\begin{equation}
f(\mathbf r, \mathbf v)d\mathbf r \,d\mathbf v \propto 
\left(-E(r,v)\right)^{7/2} r^2\, v^2\, dr\, dv
\end{equation}
where $E = \Phi(r) + \frac 12 m v^2$. At a given radius $r$, the probability of 
finding a particle with absolute velocity $v$ is given by 
\begin{equation}
\label{velocity_dist}
\tilde f (v) dv \propto \left(-E(r,v)\right)^{7/2} v^2\, dv = \left(-\Phi(r) - 
\frac 12 m v^2\right) v^2 \, dv
\end{equation}
Since the potential is known analytically at all values of $r$, the velocity 
distribution is known everywhere.

In our tests, we generate particles with density profile following 
\eqref{plummer_density} for $b=5$ in grid units. For every particle we generate 
a random velocity whose magnitude is drawn from the probability distribution 
\eqref{velocity_dist}, and whose direction is drawn isotropically. To
test advection, we  
give each particle an additional constant velocity. This constant velocity will 
have the effect of shifting the Plummer sphere in space maintaining the shape of 
the density profile. We also initialize the fluid on the grid with the same 
density profile. 

\begin{figure}[t!]
\centering
  \includegraphics[scale=0.35]{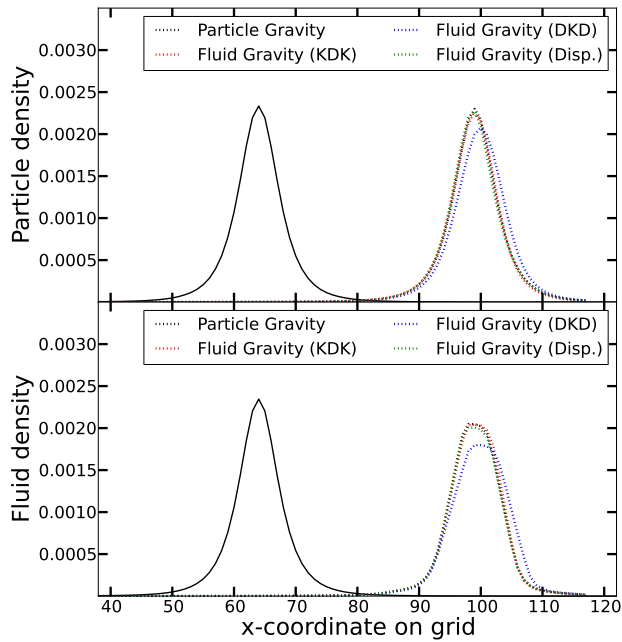}
  \caption{Density profile of an isolated advecting Plummer sphere with $b=5$. 
  The solid black lines indicate the initial density profile of particles (top panel)
  and fluid (bottom panel).
  The dotted lines in the top panel shows the final density profile of particles from 
  4 runs - a run where 
  particles source gravity (black),  a run where the fluid sources gravity and the 
  particles are evolved using the Kick-Drift-Kick method (red), a fluid-sourced 
  gravity run where the particles are evolved using the Drift-Kick-Drift method 
  (blue), and a fluid-sourced gravity run where we estimated the velocity 
  dispersion instead of the bulk velocity (green) - after $20$ dynamical times. 
  The dotted lines in the bottom panel shows the fluid profile from the same four runs after $20$ 
  dynamical times. While the other methods all agree well with the run in which particles source
  gravity, the Drift-Kick-Drift method is not suitable for coupling with 
  our hydro scheme.}
  \label{plummer1}
\end{figure}

We evolve the particles and the fluid using the methods mentioned in \S 
\ref{Methods}.  In our tests, we compare two different types of runs. In the first 
case, gravity is sourced by the particles, and so the fluid density acts as a 
tracer, evolving passively due to the motion of the particles. This provides a 
good check for our method - by comparing the particle and fluid density profiles 
at different times, we
can see if the two descriptions of the same underlying dynamics do remain 
closely coupled to each other. 

In the second type of runs, we use the fluid itself to source the gravity, which 
is the method we will use in our actual cosmological simulations. In this case, 
the particles act as tracer particles which are used only to estimate either the 
bulk velocity or the velocity dispersion on the grid. Again we check if the two 
density profiles - one from the particles and the other from the fluid - match 
each other at different times. 

We compare these runs in Fig \ref{plummer1}. We represent the particle-based 
gravity run by the black dotted lines. A fluid gravity run based on estimating 
bulk velocity from particles, and using Kick-Drift-Kick to update the particles 
is shown with the red dotted line. A similar run, but using Drift-Kick-Drift for 
the particles is plotted with the blue dotted line. Finally a fluid gravity run 
which the test particles (KDK evolution) to estimate the velocity dispersion is 
shown with the green dotted line. We see that the Kick-Drift-Kick method 
provides a better coupling to our hydro method than the Drift-Kick-Drift method. 
We also find that the run which used velocity dispersion estimation and the run 
which used bulk velocity estimation agree very well. Therefore,
whenever we do not need to smooth quantities to suppress shot noise,
we will use the faster bulk velocity estimation method.

\begin{figure}[t!]
\centering
  \includegraphics[scale=0.35]{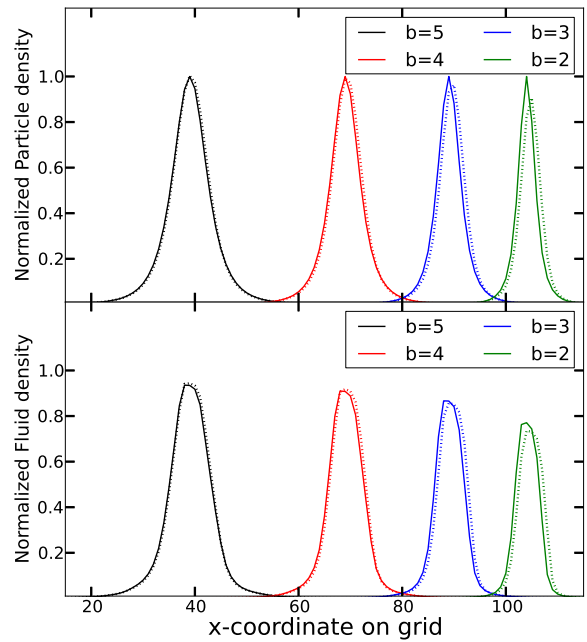}
  \caption{Density profiles for advecting Plummer spheres after $20$ dynamical 
times for $b=5$ (black), $b=4$ (red), $b=3$ (blue) and $b=2$ (green). The solid 
lines represent the profiles from runs where gravity is sourced by particles. 
The dotted lines are from runs where the fluid sources gravity. The top panel 
plots the particle density profile, normalized by the maximum value of particle 
density for a given value of $b$ from the particle based runs. The bottom panel 
plots the fluid density profiles with the same normalization. The profiles have been 
staggered for clarity.}
  \label{plummer3}
\end{figure}

We next investigate how the results from these tests are affected by the resolution 
of the advecting Plummer sphere. In the previous example with $b=5$,
the Plummer sphere was resolved by roughly ten grid elements in  
each dimension, meaning it was well resolved. In cosmological simulations, 
depending upon the shape of the power spectrum, small virialized objects which 
are not well resolved may form. Therefore, we redo our test for $b=4$, $b=3$, 
and $b=2$. We see from Fig. \ref{plummer3} that as we reduce the value of $b$, 
and therefore the resolution, the density profiles from the particle gravity 
runs after $20$ dynamical times start to diverge from the fluid
gravity runs. For the case where $b=2$, the difference in the density profile 
of particles between the two types of runs (where particles source gravity and where 
fluid soures gravity) is as much as $15\%$. These differences grow over time, and will 
cause artificial damping of small scale structures.

\begin{figure}[t!]
\centering
  \includegraphics[scale=0.35]{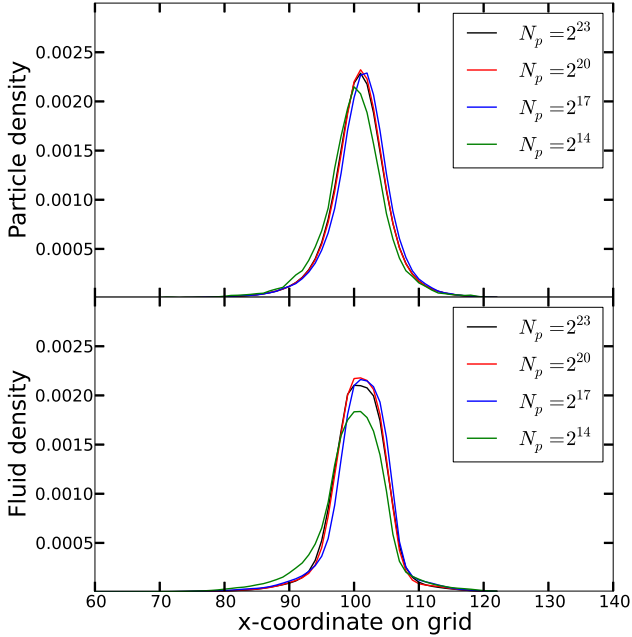}
  \caption{Particle (top panel) and fluid (bottom panel) density profiles for an 
advecting Plummer sphere with $b=5$, with different number of test particles 
($N_p$) used to sample the initial velocity distribution and estimate bulk 
velocity. We find that for this size, $N_p\sim 10^5$ is sufficient to follow the 
dynamics correctly.}
  \label{plummer5}
\end{figure}

We also test how the number of particles we use to estimate the bulk velocity or 
velocity dispersion on the grid affects the advection. We find that for $b=5$, 
the results do not change much as long as we use more than $\sim 10^5$ particles 
as seen in Fig. \ref{plummer5}.  

\subsection{Collision of Plummer spheres}
\label{plummercrash}

\begin{figure}[t!]
\centering
  \includegraphics[scale=0.37]{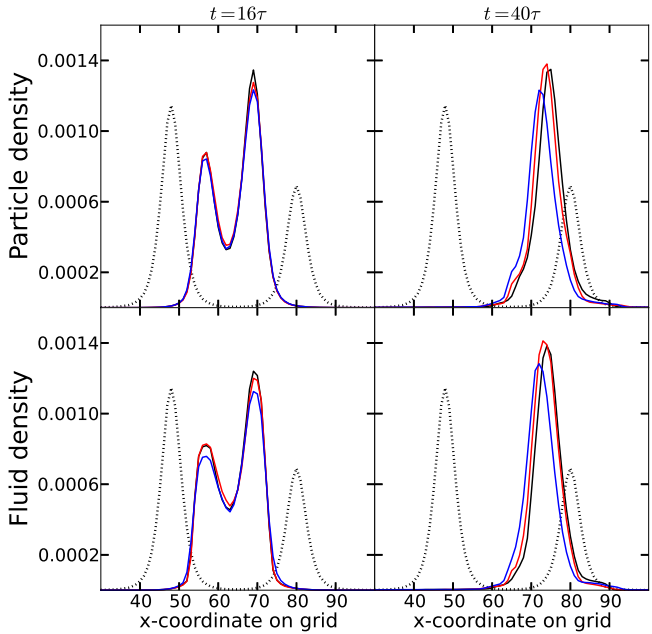}
  \caption{Evolution of the density profiles of two colliding Plummer spheres.  
The solid curves represent the particle density (upper panels) and fluid density 
(lower panels) for three different runs - gravity from particles (black), 
gravity from fluid  Superbee (SB) flux limiter (red), and another fluid gravity 
run with Monotonized Central (MC) flux limiter (blue). The dotted curve in each 
panel shows the initial configuration. The left panels represent the densities 
when the Plummer spheres have passed through each other once. The right panels 
represent densities when the two spheres have merged.}
\label{plummer2}
\end{figure}

To further test our code for dynamics of virialized objects, we next
consider what 
happens when two individual Plummer spheres are made to advect through each 
other. If the two Plummer spheres form an isolated system, the mutual 
gravitational attraction would mean that the Plummer spheres would slosh through 
each other before finally merging into one bound object. The dynamics of this 
system is analogous to the ubiquitous merger of structures one finds in typical 
cosmological simulations.

In our test, we have two Plummer spheres with the same Plummer radius ($b=5$ in 
grid units) but different masses. We use the same initialization technique as in 
the previous test - modified to take into account the different masses of the 
two spheres. We also give bulk velocities to the two Plummer spheres so that 
they move toward each other head on.

Once again, we compare what happens when gravity is sourced by the particles and 
when it is sourced by the fluid. For the latter case, we also compare how the 
results of this test is affected by our choice of the flux limiter for our hydro 
scheme. We show the comparison in Fig.\ref{plummer2}. We see that the particle 
profiles (top panels) and fluid profiles (bottom panels) of the run with the 
Superbee (SB) flux limiter (red curve) remain closer to the particle based run 
(black curve), than the run in which we used the Monotonized Central (MC) flux 
limiter \citep{Vanleer1977} (blue curves), especially at late times. This is 
understandable because the MC flux limiter is known to be more diffusive than 
the Superbee.

\subsection{Linear growth rate}

In cosmology, the high-redshift growth of the power spectrum of all
species, including light particles like neutrinos and heavy particles
like WDM, can be calculated in linear theory. 
We use the linear Boltzmann code CLASS 
\citep{Lesgourgues2011a,Blas2011,Lesgourgues2011b} to do these calculations. The 
power spectrum and transfer functions from CLASS at the starting redshift, 
$z_{\rm start}$, are used to initialize our simulations. The evolution of the power 
spectrum from the simulation boxes can then be compared at later redshifts to 
the outputs from CLASS at those redshifts.

\begin{figure}[t!]
\centering
  \includegraphics[scale=0.4]{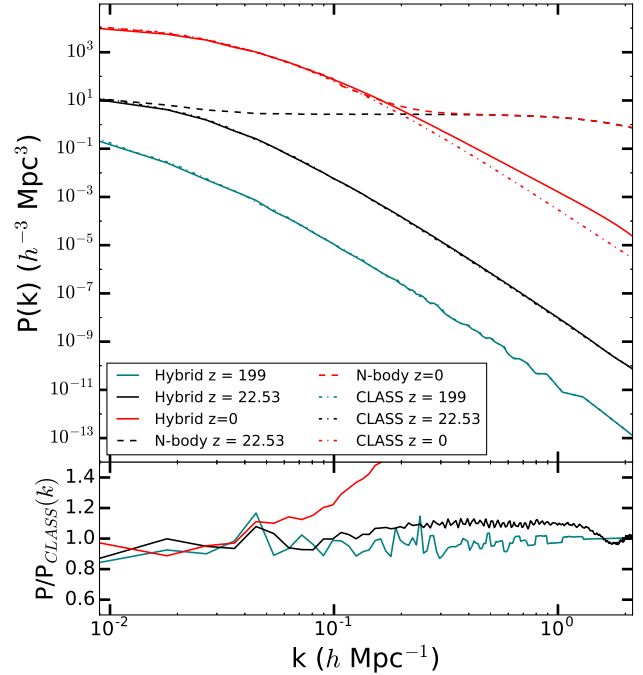}
  \caption{Comparison of the growth of the power spectrum of neutrinos 
($m_\nu=0.1\,$eV). The top panel plots the comparison of the power spectra from 
the hybrid simulations (sloid lines) to the outputs from CLASS (dot-dashed lines)
. We plot the initial power 
spectra at $z=199$, as well as at redshifts of $z=22.53$ and $z=0$. 
We also plot the power spectra (dashed lines) at the later redshifts from 
an N-body treatment of neutrinos, which are dominated by shot noise at small 
scales. In the bottom panel we plot the ratio of the power spectra from the 
hybrid simulations to the linear theory results at $z=199$ (teal), $z=23.53$ 
(black) and $z=0$ (red).}
 \label{nugrowth}
\end{figure}

In simulations involving light neutrinos, linear perturbation theory can be used 
to describe the evolution of the neutrino power spectrum for most of the 
evolution, except maybe at very late times. This means that for most of the 
simulation, the smoothing length defined in Eqn.\ \eqref{smoothing_length} is larger 
than a pixel, and we need to estimate the velocity dispersion from the test 
particles. Comparing the growth of the power spectrum at these times against 
CLASS provides a test for the code when it is solving both Eqns.\ \eqref{continuity} and 
\eqref{euler}. Note that at these early times, the growth of the
neutrino power spectrum is scale dependent.  On scales 
larger than the free streaming scale of the neutrinos, the neutrino
power spectrum grows at the same rate as the CDM power spectrum.
Below the free streaming scale,  however, different scales can grow at
different rates. To compute the growth factor accurately on all scales, 
it is essential to determine the correct ``effective sound 
speed'' for the neutrinos. This is exactly what we estimate from the velocity 
dispersion. The effective sound speed depends on redshift, meaning
that a comparison of the power spectrum at different times provides a test for the 
accuracy of our estimates of the velocity dispersion. In Fig.\ \ref{nugrowth}, we 
show the growth of the neutrino power spectrum for a neutrino species of mass 
$m_\nu=0.1\,$eV at redshifts $z=22.53$ and $z=0$. We compare the results 
from our simulations to the growth predicted by CLASS for the same cosmology. We 
see that the power spectra from our simulations match the linear theory 
predictions quite well at all scales in the simulation box at $z=22.53$. Leaving aside the 
large scales, which are affected by sample variance, there are, however, 
differences of about $10\%$ even on the smaller scales. This arises from the 
fact we use the particles directly to get the velocity dispersion, while in CLASS, an approximation is used to evaluate the sound speed when the fluid approximation is turned on \citep{Lesgourgues2011b}. We have checked that if we use the same approximation in our code, instead of the stress tensor from the particles, we match the CLASS results to within a few percent. However, this small difference in the neutrino power 
spectrum will have minimal effect on the observable matter power spectrum. We note that all our comparisons with CLASS were with the fluid approximation turned on, which means that the effective sound speed of the neutrinos is treated to be scale independent. While this is done explicitly in CLASS when the fluid approximation is turned on, in our simulations, the large smoothing length at early times for light neutrino species means that the velocity dispersion is scale independent over much of the simulation box - effectively using the fluid approximation. 

To illustrate the effect of shot noise in neutrino simulations, we
also plot in Fig.\ \ref{nugrowth} the neutrino power spectrum at the 
later redshifts. These spectra were obtained from a simulation which treats the
neutrinos as N-body particles. The power spectrum of neutrinos from  
this simulation is dominated by shot noise for most scales in the box - this is true even at 
$z=0$.

\begin{figure}[t!]
\centering
  \includegraphics[scale=0.4]{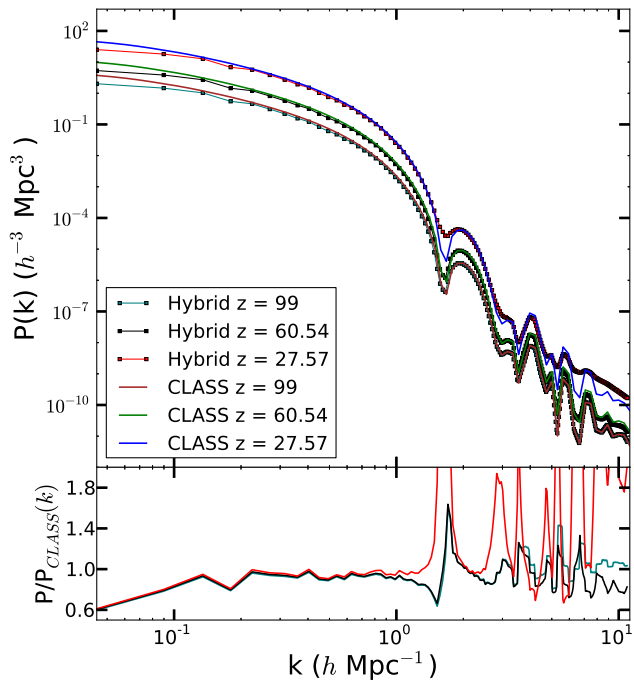}
  \caption{Linear growth of the power spectrum for a WDM particle of mass 
$200$eV. The top panel plots the power spectrum at initial redshift $z=99$ and 
at $z=60.54$ and $z=27.57$ from the hybrid simulations and from linear theory. 
The bottom panel plots the ratio of the power spectra from the simulations and 
from linear theory at $z=99$ (teal), $z = 60.54$ (black) and $z=27.57$ (red). At 
large scales in the box, the ratio is different from $1$ due to sample 
variance.}
  \label{wdmgrowth}
\end{figure}

In Fig.\ \ref{wdmgrowth}, we show an example of a WDM simulation.  In
this case, while the power spectrum is initially damped on small 
scales, non-linear structure does develop at high $k$ as time
progresses. The power spectrum on these scales will then disagree with
the linear CLASS power spectrum. However, if the 
simulation volume is large enough, then the largest scales in the box will still 
be well described by linear theory, and can be matched to the outputs from 
CLASS. Unlike the neutrino simulations, where the energy density and 
gravitational potential are dominated by CDM particles, the 
growth of the matter power spectrum in the WDM simulations is governed by the 
fluid, and testing the growth rate in these simulations provides a stronger test 
of the coupling between the fluid equations and the gravitational potential. 
Fig.\ \ref{wdmgrowth} shows the growth of the power spectrum from a $200$eV WDM 
particle compared to the outputs of CLASS at two different redshifts, $z=60.54$ 
and $z=27.57$. At these early times, our method is able to reproduce the linear 
results down to almost the smallest scales in the box.

\subsection{CDM comparison run}
\label{cdmcomp}

Even though our method is meant to be used in cosmologies which include 
particles whose thermal velocities cannot be ignored, we can test the validity 
of our method by using it for a standard CDM cosmology run, and comparing the 
results to those of an N-body simulation. Since N-body simulations are known to 
be accurate for CDM-only cosmologies, this comparison allows us to test our 
method starting at high redshifts when linear theory is valid to late times and 
low redshifts where we can test the non-linear evolution of our method.

For this comparison, we do not need to smooth quantities, since there are no 
random particle velocities which will lead to thermal shot noise. We use the 
test particles to measure the bulk velocities on a grid at all redshifts, and 
use these velocities to solve the continuity equation \eqref{continuity} to 
evolve the densities forward. Both simulations are done on a $140\,h^{-1}$Mpc 
box with $\Omega_\Lambda = 0.7$ and $\Omega_{\rm CDM}=0.3$. The Hubble
constant $H_0$ is taken to be 
$70\,$km/s/Mpc. Both sets of simulations used $512^3$ particles with a $512^3$ 
grid to calculate the gravitational potential. Our method also uses a $512^3$ 
grid to store densities and bulk velocities. Initial conditions are generated at 
redshift $z=99$ using CLASS.

\begin{figure}[t!]
\centering
  \includegraphics[scale=0.4]{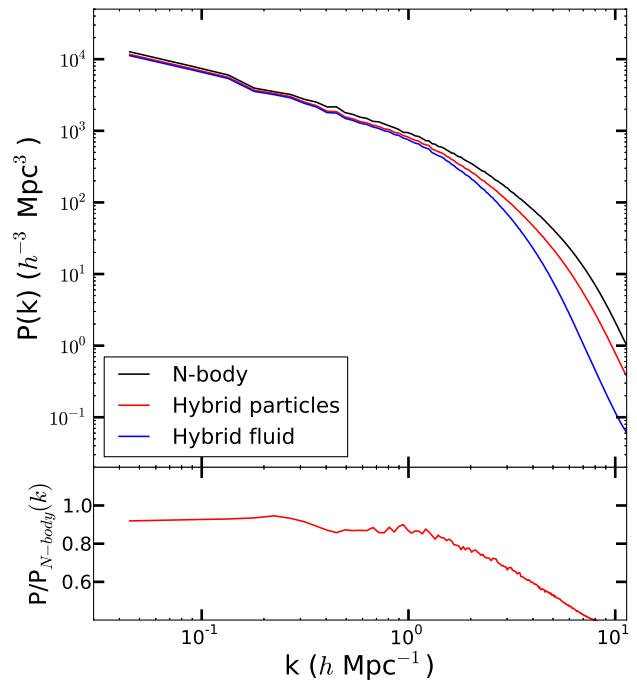}
  \caption{Final power spectrum at $z=0$ for CDM runs. In the top panel, we plot 
the results a normal N-body run (black) as well as from the particles (red) and 
fluid (blue) in our hybrid simulations. In the bottom panel we plot the ratio of 
the particle power spectrum from the hybrid simulations to the power spectrum 
from the N-body run.}
  \label{cdmps}
\end{figure}

We begin by comparing the growth of the power spectrum on scales which remain 
linear - the largest scales in the simulation box. At early times, the growth at 
these scales matches the growth from standard N-body simulations. This agreement 
persists even to times when halos start forming in the box. However, at very 
late times, the agreement breaks down, and the largest scales grow at a rate 
slower than in N-body simulations. This can be seen clearly from the lower panel 
in Fig. \ref{cdmps}, where the ratio of the power spectrum from our hybrid 
simulation to the power spectrum from the N body calculation (both at $z=0$) is 
less than $1$ at even the largest scales in the box.

At late times in these CDM simulations, structures form at all scales and at all 
locations. Even deep inside voids, there are small fluctuations in the density 
field. Because of these small variations, the density field is not 
smooth over most of box, and there are large numbers of local extrema
and saddle points along any of the axes of the simulation box. As
mentioned in \S \ref{hydromethod}, our  
hydrodynamic scheme needs to switch over to a spatially first order scheme 
whenever it encounters a saddle point - unphysical oscillations set in when this 
condition is not satisfied. Due to the numerous saddle points which develop late 
in these CDM simulations, our method is forced to solve the governing equations 
of motion in a spatially first order manner over most of the pixels in the box. 
First order methods are known to be highly diffusive - even up to scales 
comparable to the entire simulation box. Because of this, the power at large 
scales is damped, and the growth rate of these scales becomes
unphysically slow.  We found that most of the pixels at which 
the hydrodynamic method is forced to go first order lie inside voids and 
regions where $|\delta| \sim 1$, rather than inside halos or regions where 
$|\delta| \gg 1$. Note that
this problem is not as severe for WDM cosmologies in which the streaming 
scale is resolved. However, there is still numerical diffusion on small scales 
compared to an N-body treatment, and we will discuss below in
\S\ref{sec:wdm} how this can be remedied using an SPH-like approach at
early times.

\begin{figure}[t!]
\centering
  \includegraphics[scale=0.4]{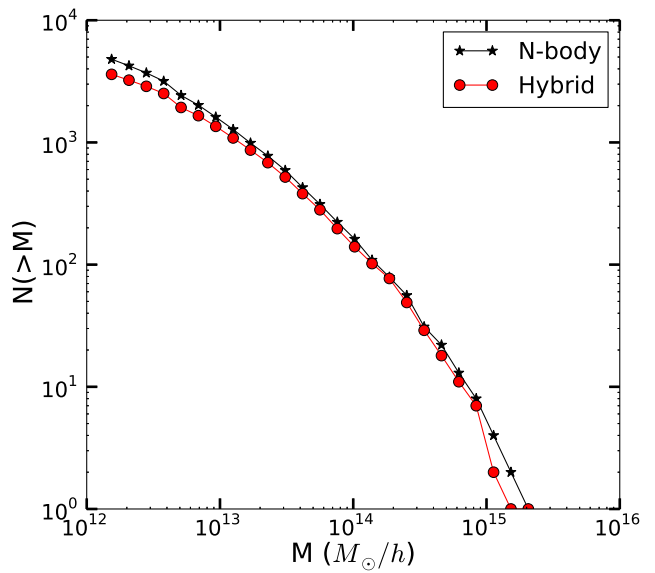}
  \caption{Mass function from the N-body run (black) and the hybrid Eulerian method 
(red) for a $\Lambda$CDM cosmology. The red curve shows a deficit of halos at all mass scales,
including near the high end of the mass function.  In contrast, the hybrid
Lagrangian (SPH) method is formally identical to N-body for CDM cosmologies.}
   \label{fig:cdmmassfunc} 
\end{figure}

To further illustrate the effects of artificial diffusion, we next compare the mass 
functions of halos from our simulations to N-body halo mass
functions. To find halos in our simulations, we use the Rockstar halo
finder \citep{Behroozi2012} as well as a spherical overdensity halo
finder. For the latter method, we define a halo as  spherical regions
of radius $R_{\rm halo}$ around overdensity peaks within which the average
enclosed matter overdensity is greater than $200$. The associated halo
mass is then defined as 
\begin{equation}
M_{\rm halo} = \frac 4 3 \pi R_{\rm halo}^3 \times \left(200 \, \Omega_m 
\rho_{\rm crit}\right)
\end{equation}
where $\rho_{\rm crit}$ is the critical density of the universe.

In our simulations we can use either the fluid density or the test particle 
density to define halos. Since there is no thermal shot noise in the test 
particle density field in these CDM simulations, there is no problem using the 
test particle densities in our halo finder. However, we can also do the same 
even in WDM simulations where there is shot noise from the thermal velocities of 
particles. This is because throughout the simulation, we never use the test 
particle density to source gravity, hence the shot noise in the density field is 
not allowed to grow gravitationally. Therefore the shot noise fluctuations in 
the density field of the test particles arise at an early time and become 
frozen once the particles cool down. These initial fluctuations are much smaller than  
overdensities of $\delta \sim 200$ that are needed for a halo to be detected. 
Therefore, the halos that we detect at late times in the density field of test 
particles are real halos and not shot noise artifacts. This would not have been 
true if we had used the test particle density instead of the fluid density in 
the Poisson equation - in that case, the small shot noise fluctuations would 
have grown gravitationally over time and produced spurious halos.

Comparing mass functions, we find that if the halos defined using the test 
particle densities rather than fluid densities, the results match those from the 
N-body simulations more closely. This is expected from the tests we performed in 
\S \ref{plummerad} and \S \ref{plummercrash} - the fluid profile near the 
center is flattened by artificial diffusion. The particle profile is not as 
affected on scales which are well-resolved on the grid. However, there are still 
differences between the mass functions from our simulations compared to the 
N-body mass functions. Also, there are differences on small scales between the 
N-body power spectrum and the power spectrum from our simulations. This is once 
again due to artificial diffusion on small scales. For small halos with steep 
profiles, the flattening of the fluid profile means a large fraction of the mass 
is moved away from the peak. This also ends up affecting the test particles 
which respond to the gravitational field of the fluid.

\begin{figure*}[t!]
\centering
  \includegraphics[trim={0 3cm 0 3cm},scale=0.7]{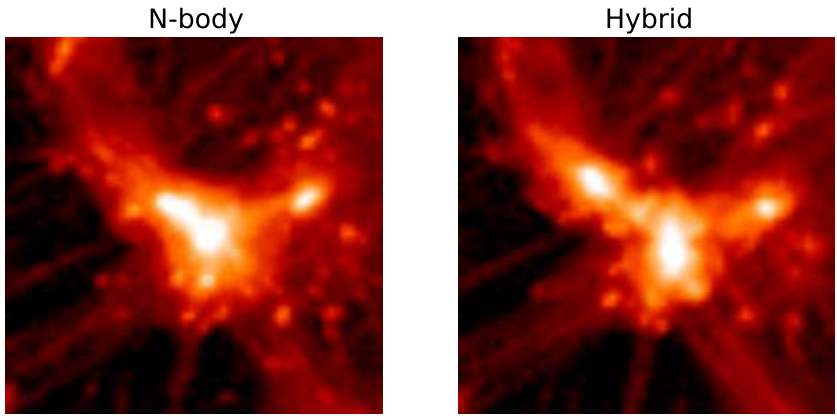}
  \caption{A zoom-in of a halo at $z=0$ from the CDM comparison simulations - 
the left panel shows the density field from a standard N-body simulation, while 
the right panel shows the density field from our hybrid method. In the N-body 
case, the central object has already merged, whereas in the hybrid method, the 
two objects are still distinct.}
    \label{fig:cdmhalo}
\end{figure*}

This loss of smaller scale power due to diffusion leads to fewer smaller halos 
found in our simulation box than in the corresponding N-body boxes. 
However, at a given redshift, this can also cause differences in the masses of 
large, fast-accreting halos. This is because the lack of small scale power 
alters the dynamics of the halos and also the time at which mergers happen 
- the mergers in our simulations lag behind the corresponding merger in the 
N-body simulation, as can be seen in Fig.\ref{fig:cdmhalo}. The fast acrreting 
halos at a given redshift have a rapid change in their mass because of 
successive mergers. Because these mergers have not yet happened in the our 
simulations, the halos are at a lower mass. This shows up in the mass function 
as a lack of halos at the largest masses, as can be seen in Fig.\ 
\ref{fig:cdmmassfunc}.

Our findings show that in the CDM case, numerical diffusion in our grid-based 
fluid method affects all useful quantities measured from the simulation boxes - 
even at the largest scales. This motivates us to use a more Lagrangian approach 
for simulations in which there is structure on all scales in the box.
Specifically, we adopt the SPH-like approach that we outlined in
\S \ref{sphmethod} for WDM simulations where the fluid that we are
simulating is the main source of  the gravitational potential.  As the
results in this section have shown, for such simulations any numerical
diffusion becomes important.  In contrast, for neutrino simulations, where the gravitational 
potentials are highly dominated by the CDM, we will continue to use grid-based 
Eulerian methods, as our implementation of SPH is computationally more 
expensive than our implementation of grid-based hydrodynamics.

\subsection{Simulating Warm Dark Matter cosmologies}
\label{sec:wdm}

In our final test, we simulate WDM cosmologies to see if our fluid
methods can eliminate the artifacts generated for these cosmologies when 
simulated using N-body techniques. For N-body simulations which include the 
random thermal velocities of the WDM particles - or a hot start - these 
artifacts are spurious halos seeded by shot noise in the density field. For cold 
start simulations which do not include the thermal velocities, the artifacts are 
``beads on a string'' halos studied extensively by \citet{Wang2007}.

For WDM, there is a characteristic free streaming scale below which the linear 
power spectrum is damped. Above this scale, the behavior of WDM is 
the same as CDM, so we will concentrate on simulations in which the 
free streaming scale is resolved. To estimate the scale of this damping, we take 
the linear power spetrum at $z=0$ for CDM and the WDM particle we are interested 
in. We find the $R_{\rm damp}$ such that the CDM power spectrum convolved with 
$\left|W(kR_{\rm damp})\right|^2$ gives the WDM power spectrum 
\citep{Villaescusa2011b}, where $W(kR)$ is the Fourier transform of the top-hat 
window function with radius $R$. We also define a damping mass scale 
\begin{equation}
\label{dampingmass}
M_{\rm damp} = \frac 4 3 \pi R_{\rm damp}^3 \Omega_m \rho_{\rm crit}
\end{equation}
where $\rho_{\rm crit}$ is the critical density of the universe.

For these WDM simulations, we use the hybrid SPH method laid out in 
Sec.\ \ref{sphmethod} to suppress the effects of numerical diffusion. It is 
important to do so, because we are interested in determining the halo mass 
function below the damping scale down to the smallest scales in the box, and 
these scales are most affected by diffusion.

To validate our hybrid SPH method, we first run a simulation for a WDM particle 
with mass $m=200 \, $eV, for which the damping scale is
$3.3 h^{-1}\,$Mpc. We do three runs for this cosmology, a hot start
N-body run, a cold start N-body run, and a run using our hybrid SPH
method. The initial thermal velocities are included in both the hot
start run and our SPH method. These simulations  were done on $512^3$
grids with $1024^3$ particles, and for the background  cosmology we
used $\Omega_{\Lambda} = 0.7$, $\Omega_m = 0.3$ and
$H_0=70\,$km/s/Mpc.  The size of the simulation volume 
was $\left(140 h^{-1}{\rm Mpc}\right)^3$.

\begin{figure}[t!]
\centering
  \includegraphics[scale=0.4]{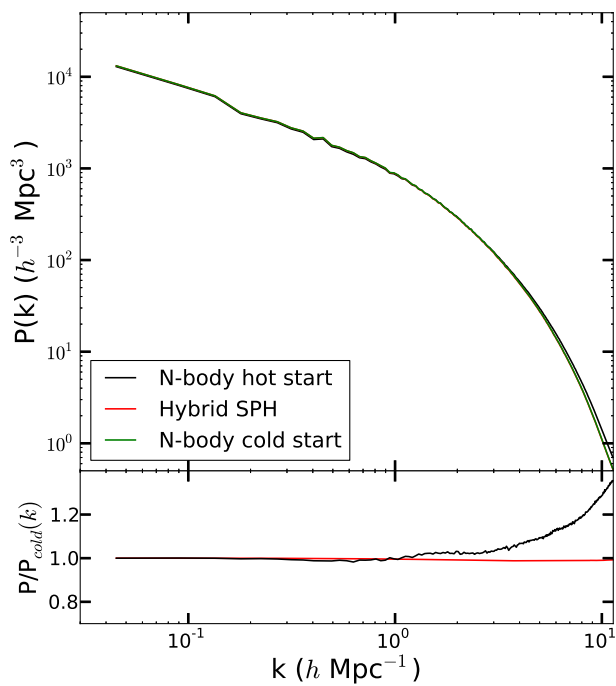}
  \caption{Final power spectrum at $z=0$ for WDM species with mass $200\,$eV. In 
the top panel, we plot the results from a hot N-body run (black), a cold N-body 
run (green), as well as our hybrid SPH simulation (red). The results from the 
SPH run and the cold start N-body run are virtually indistinguishable. On large 
scales in the box, the hot start N-body run agrees with the other two, but there 
are differences on small scales. In the bottom panel we plot the ratio of the 
particle power spectrum from the hybrid SPH run to the cold N-body power 
spectrum (red) to show that these methods match each other down to the smallest 
scales resolved by this simulation box, even though the initial conditions for 
the SPH run was the same as that for the hot start N-body run. We also plot the 
ratio of the power spectra from the hot run to the cold run (black) to show 
their difference on scales affected by the streaming of particles.}
  \label{wdmps}
\end{figure}

We first compare the final power spectra from all three simulations at $z=0$, 
as illustrated in Fig.\ \ref{wdmps}. We see that at large scales, the power 
spectra from all three runs agree with each other. However, at small scales, 
while the cold start N-body run and our hybrid SPH run give the same results, 
the power spectrum from the hot run starts to deviate from the others. This is 
expected in the hot run as the shot noise due to random streaming of the 
simulation particles leads to the presence of extra power at small scales, and 
this noise grows over the course of the simulation. However, we note that even 
though we included these thermal velocities in our hybrid SPH method, the final 
result agrees very well with that from the cold start run, down to the smallest 
scales that we resolve in this simulation box. This can be seen clearly from the 
bottom panel of Fig.\ \ref{wdmps}, where we plot the ratio of the two power 
spectra, and find that the ratio is very close to $1$ at all scales. 

\begin{figure}[t!]
\centering
  \includegraphics[scale=0.4]{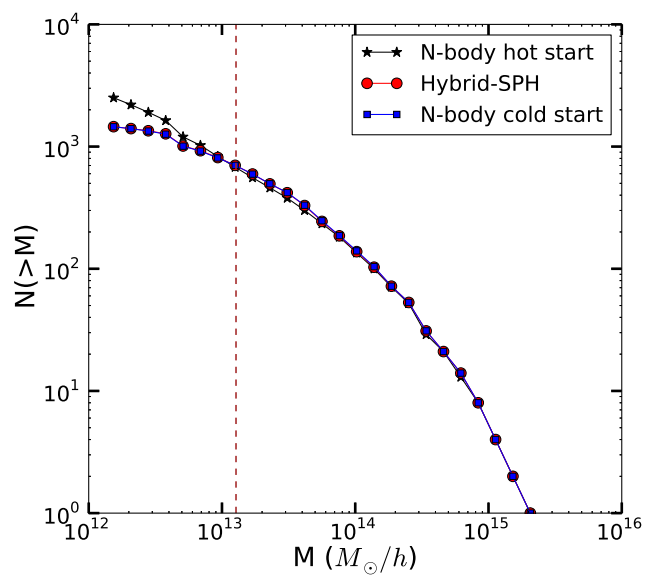}
  \caption{Mass function for WDM species with $m=200\,$eV. We plot results from a 
hot N-body run (black), a cold N-body run (blue), and our hybrid SPH method (red). 
The brown dashed line represents $M_{\rm damp}$ defined in Eqn.\ \eqref{dampingmass}. The 
hot run shows the presence of many halos below $M_{\rm  damp}$, while the other runs 
do not show these smaller halos.}
  \label{massfunction}
\end{figure}

We also compare the halo mass functions from the three runs in Fig.\ 
\ref{massfunction}. We expect the effect of the streaming length of the WDM 
particle to show up at masses below $M_{\rm damp}$ represented in the plot by the 
dashed brown line. Since the power spectrum was initially damped on these 
scales, we expect very few halos with these low masses to form in the box. That 
seems to agree with what we see from both the cold start run and the hybrid SPH 
run - the cumulative mass function flattens off at mass scales smaller than 
$M_{\rm damp}$. However, the hot run does show the presence of many small halos. 
These are the spurious halos which were seeded by the shot noise arising from 
the thermal motions at early times. 

These results suggest that our hybrid SPH method is effective in eliminating the 
effects of shot noise.  Even though we started off by taking into account the 
thermal velocities of the particles from the Fermi-Dirac distribution, our final 
results agree very well with simulations which do not include these thermal 
velocities, and are therefore immune to this form of shot noise.

However, as discussed earlier, the cold start simulations for WDM 
show ``beads on a string'' artifacts \citep{Wang2007}, produced when structures 
collapse along grid lines in the simulations. We check if our method can be used 
to get rid of these artifacts, which once again shows up in the mass functions 
at scales much smaller than the damping scale. For this, we simulate a lighter 
WDM particle, $M=60 \,$eV, for which $R_{\rm damp}\sim 10.2\, h^{-1}$Mpc. This 
allows us to resolve scales much smaller than the streaming scale in our 
simulation box. We choose our backgorund cosmology similar to that used in 
\citet{Wang2007}: $\Omega_m = 1$, $H_0 = 70 \,$km/s/Mpc and $A_s = 4.6 \times 
10^9$, where $A_s$ is the amplitude of the primordial power spectrum. The size 
of the simulation volume was $\left(70 h^{-1}{\rm Mpc}\right)^3$. As before, we run 
both hot start and cold start N-body runs to compare to our simulations. 

\begin{figure}[t!]
\centering
  \includegraphics[scale=0.4]{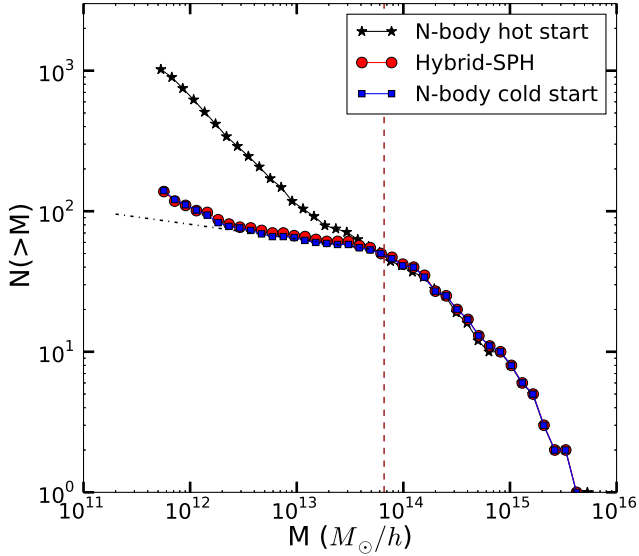}
  \caption{Mass function for WDM species with $m=60$eV. We plot results from a 
hot N-body run (black), a cold N-body run (blue), and our hybrid SPH method (red). 
The brown dashed line represents $M_{\rm damp}$. Below $M_{\rm damp}$ the hot run shows 
a number of spurious halos. At even smaller scales, both the cold start and the 
hybrid SPH run start showing ``beads on a string'' artifacts. The black 
dash-dotted line is an extrapolation from the flat part of the halo mass 
function.}
  \label{massfunction2}
\end{figure}

We compare the mass function from the three runs in Fig. \ref{massfunction2}. 
Once again we see that the hot run produces many spurious halos below 
$M_{\rm damp}$. In the cold run and the hybrid SPH run, we see that the cumulative 
mass function flattens just below the damping scale, meaning that very few halos 
are produced in this mass range. At scales much smaller than the damping scale, 
the cold start run shows an up-turn in the mass function. This upturn is due to 
the halos collapsing along grid lines in simulations because of very low power 
on small scales. The black dash-dotted line is an extrapolation of the flat part 
of the mass function to show how these ``beads on a string'' halos affect the 
mass function. We see the same feature even in the hybrid SPH run, where the 
mass function follows that from the cold start N-body closely down to the scales 
where the artificial halos start showing up.

From this test we conclude that while our hybrid method is highly effective in 
eliminating shot noise in WDM simulations where the initial thermal velocities of 
particles are taken into account, they cannot get rid of the ``beads on a 
string'' artifacts seen in cold start N-body simulations of WDM 
cosmologies. Since this hybrid method is computationally much more expensive 
than N-body methods, and since applying our method to WDM simulations yields no 
advantage over the traditional cold-start N-body simulations, we do not find a 
persuasive reason to use this in further studies of WDM cosmologies.

\section{Neutrino Simulations}
\label{nusim}

In this section, we discuss examples where our simulations are applied to 
cosmologies containing massive neutrino species. We will mostly focus on
relatively low-resolution simulations of large volumes that contain a large
fraction of the neutrino free streaming scale; in future work we will study
nonlinear structure formation with neutrinos at higher resolution.
Because of the low spatial resolution, most halos will be unresolved.
However, cosmic voids can be 
studied using these simulations, and as we show below, massive
neutrinos can have interesting effects on the large-scale clustering
of highly empty voids.  Indeed, from simple physical arguments we can
expect that neutrinos will have a more significant impact on voids
than on halos.   While some neutrinos, mostly coming from the
low momentum tail of the initial distribution function, do get
captured by massive halos at late times, the average enclosed
overdensity $\Delta_\nu$ of neutrinos is still much smaller than the
average enclosed overdensity of CDM, $\Delta_{\rm CDM}\sim 200$. 
Any effects from neutrinos would be expectd to be of the 
order of $f_\nu \Delta_\nu/ \big(f_{\rm CDM}\Delta_{\rm CDM}\big) \lesssim 1\%$. This 
effect has been verified already \citep{Villaescusa2014,Loverde2014,Loverde2016}. Inside 
deep voids, on the other hand, the CDM overdensities are $\sim -1$, while the 
neutrino overdensities are not as negative due to their thermal dispersion. This 
means that the overall mass of the neutrinos in the void can be comparable to 
the total mass of CDM. Therefore we might expect any effects due to
neutrinos to be magnified for voids, compared to halos.
Note that throughout this section, we define voids using a threshold
on the total matter density, including both CDM and neutrinos. This is in 
contrast to voids defined using only the CDM densities, as used  
by \citet{Massara2015} to study the profiles of voids in the presence 
of neutrinos.

\begin{figure*}[t!]
\centering
  \includegraphics[trim={2cm 5cm 2cm 5cm}]{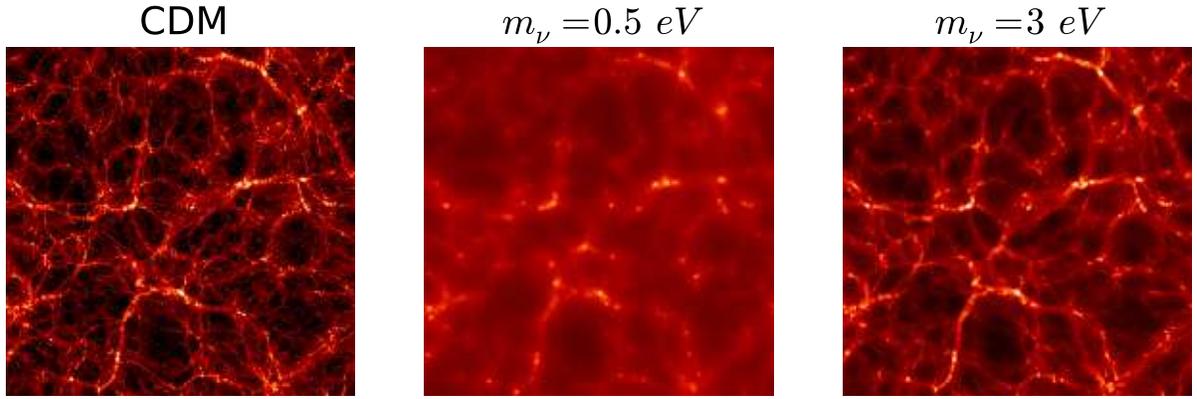}
  \caption{A slice through the density field in the simulation boxes at $z=0$ 
for two neutrino masses - $0.5\,$eV and $3\,$eV. The simulation box
side length is $175\, h^{-1}\,$ Mpc. The left panel shows the CDM
density field, the middle panel shows the density field for the 
$0.5 \,$eV neutrino, while the right panel shows the density field for
the $3\,$eV neutrino. For the lighter neutrino, on large scales, the
density traces the underlying cosmic web structure laid down by the
CDM component, but is much more diffuse on small scales. For the
heavier neutrino, the density field is less diffuse and follows  
the CDM density more closely down to smaller scales.}
    \label{fig:nucomp}
\end{figure*}

\begin{figure}[]
 \centering
 \includegraphics[scale = 0.4]{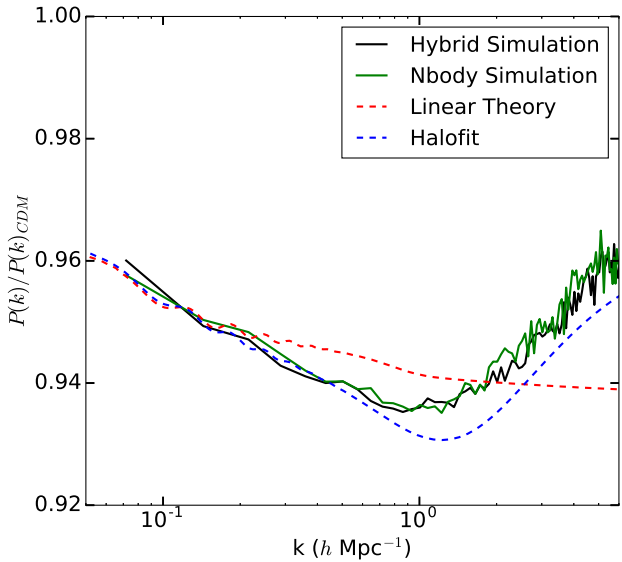}
 \caption{Relative damping of the matter power spectrum in the presence 
 of a massive neutrino with $m=0.1 \,$eV with realistic energy density ($\Omega_\nu \approx 0.0022$),
 compared to the CDM-only prediction. The linear theory prediction is 
 plotted with the dashed red 
 while the dotted blue curve represents the HALOFIT predcition. The solid
 black curve is the result of our hybrid simulation, while the solid green 
 curve is from a simulation treating neutrinos as another set of N-body
 particles. \label{fig:nonlin_damp}}
\end{figure}

Since our purpose here is to demonstrate and highlight certain effects, 
for the void studies, we will 
use parameters which help best illustrate these. Our simulations use a 
box whose size is $700h^{-1}$Mpc with $\Omega_\Lambda = 0.7$ and
$\Omega_{\rm CDM} = 0.27$, along with a single 
neutrino species of mass $m_\nu = 0.1\,$eV.  Neutrinos of this mass
would give energy density $\Omega_\nu < 10^{-3}$, producing relatively
subtle effects on nonlinear structure formation.  To amplify neutrino
effects in our simulations, we therefore use an unphysically large
number density, to give $\Omega_\nu=0.03$.  We stress that this is
only for illustrative purposes; later on we will show results for
realistic energy densities.  For this $\Omega_\nu$, we have
\begin{equation}
\label{fnu}
f_\nu = \frac{\Omega_\nu}{\Omega_{\rm CDM}+\Omega_\nu} = 0.1.
\end{equation}
The Hubble constant $H_0$ is taken to be $70\,$km/sec/Mpc. We use
$512^3$ CDM particles and $1024^3$ neutrino test particles with a
$512^3$ grid for hydrodynamic quantities. Initial  conditions for both
species are generated using CLASS at redshift $z=49$. 

Non-linear structures like voids are known to be biased tracers of the 
underlying matter field. The overdensity in the void field is denoted by 
$\delta_{\rm void}$. If the universe contains only one species, say CDM, we can 
write
\begin{equation}
\label{deltavoid}
\delta_{\rm void}(k) = b(k) \delta_{\rm CDM}(k)
\end{equation}
where $b(k)$ is the scale-dependent bias factor. On large 
linear scales, $b(k)$ is independent of $k$ for these CDM only
simulations \citep{Scherrer1998}.   
Using Eqn.\ \eqref{deltavoid}, this linear (large-scale) bias factor is given by 
\begin{equation}
b = \frac{P_{\rm void,CDM}(k)}{P_{\rm CDM}(k)}
\end{equation}
where $P_{\rm void,CDM}(k) = \left\langle \delta_{\rm void}^*(k) 
\delta_{\rm CDM}(k)\right\rangle $ is the cross spectrum between the void 
overdensity field and the underlying CDM overdensity field and 
$P_{\rm CDM}(k)$ is 
the CDM auto-power spectrum.

\begin{figure}[t!]
\centering
  \includegraphics[scale=0.4]{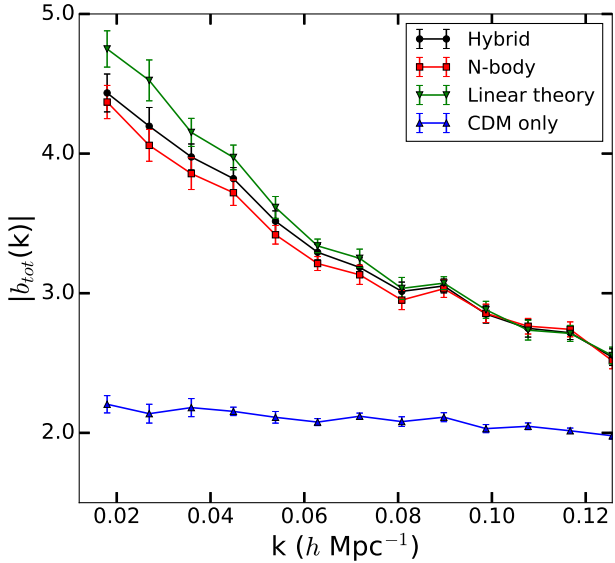}
  \caption{Absolute value of the bias defined in Eq. \eqref{totalbias} as a 
function of scale, averaged from 8 runs for $m_\nu = 0.1 \,$eV and $f_\nu = 0.1$. The voids were defined 
using the underdensity in the total matter field. The threshold for void definition 
was set at $-0.7$. We compare the results from our 
simulations (black) to the other existing methods of treating the neutrinos as a 
linear fluid (green) and treating neutrinos as a set of particles with a 
different mass in N-body simulations (red). Our method and the N-body method 
yield results which match to within error bars, but show a strong scale 
dependence. The linear method shows an even stronger scale dependence. We also 
plot (in blue) the bias for voids defined in exactly the same manner from 8 CDM-only runs in for 
which the final power spectrum matches the final CDM power spectrum for the runs 
including neutrinos.}
  \label{bias1}
\end{figure}

If a new species is added to the matter content of the universe, biasing
is no longer as simple. In this case, the void overdensity depends on both the 
CDM and neutrino overdensities, $\delta_{\rm CDM}$ and $\delta_\nu$. Eqn.\ 
\eqref{deltavoid} should then be replaced by
\begin{equation}
\label{scaledepbias}
\delta_{\rm void}(k) = b_{\rm CDM}(k) \delta_{\rm CDM}(k) + b_\nu(k) \delta_\nu(k)
\end{equation}
In terms of the total underlying matter field 
\begin{equation}
\label{deltatot}
\delta_{\rm tot} = f_{\rm CDM}\delta_{\rm CDM}+f_\nu \delta_\nu
\end{equation}
we can write
\begin{equation}
\label{totalbias}
\delta_{\rm void}(k) = b_{\rm tot}(k) \delta_{\rm tot}(k)
\end{equation}
Even on large scales where the perturbations are linear, the neutrino power 
spectrum and the CDM power spectrum can be different from one another,
implying that $b_{\rm tot}(k)$ will not be scale independent. This
scale dependent bias in large-scale structure can be an extremely
powerful signature because it does not arise in standard
cosmologies. Producing scale dependent bias on linear scales generally
requires violating locality of the formation of biased tracers, for
example due to non-gaussianity \citep{Dalal2008} or large scale
modifications to  gravity \citep{Jain2010}.

In the case of massive neutrinos, the scale dependence of $b_{\rm tot}(k)$ depends 
on the difference of the two power spectra as a function of scale, and hence the 
mass of the neutrino species. For heavier neutrinos, the free streaming scale 
may lie well inside the simulation box. We know that on scales well above the 
free streaming scale, perturbations to both neutrinos and CDM evolve similarly, 
and so their power spectra on large scales will be identical at late times. In 
this case $b_{\rm tot}(k)$ will indeed be scale independent. However for lighter 
neutrinos with very long free streaming lengths, the power spetcra at linear 
scales will be very different for the neutrinos, compared to CDM. In this case, 
$b_{\rm tot}(k)$ can be strongly scale dependent.

In our simulations we define the total matter overdensity as in  Eqn.\ 
\eqref{deltatot} and then use spherical overdensity void finder. We define voids 
as those regions around a density minimum in which the averaged overdensity is 
below the cutoff of $-0.7$. We then select the largest voids in the box, so that 
we are not affected by exclusion effects. We run 8 sets of simulations with 
different realizations of the power spectrum. We calculate the linear bias in 
each of these 8 realizations, and then average over them to reduce the sample 
variance.

\begin{figure}[t!]
\centering
  \includegraphics[scale=0.4]{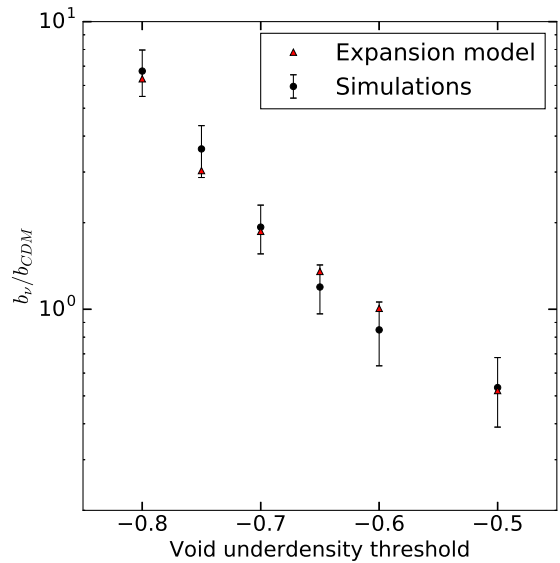}
  \caption{Comparison of the ratio of Eulerian biases $b_\nu$ and $b_{\rm CDM}$ at $f_\nu = 0.1$,
  defined in Eqn.\ \eqref{scaledepbias}, measured in simulations
  (black circles) vs.\ predictions from the spherical expansion model
  (red triangles), as a function of the threshold floor used in the
  void definition. The bias ratio diminishes for decreasing void
  thresholds, a trend quantitatively predicted by the spherical
  expansion model. The error bars represent the errors on the best fit values from the 
  simulations.}
  \label{model}
\end{figure}

\begin{figure}[t!]
\centering
  \includegraphics[scale=0.4]{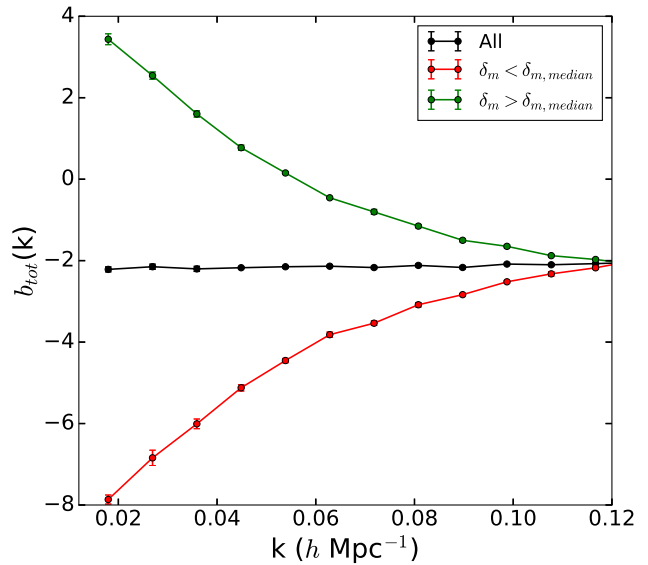}
  \caption{Behavior of the void bias from the hybrid simulations when
    the voids are selected using the CDM field only, and then split on
    the basis of the enclosed total matter overerdensity for $m_\nu =
    0.1\,$eV and $f_\nu = 0.1$. The black curve represents $b_{\rm
      tot}(k)$ for voids selected using the CDM field only with
    overdensity threshold of $-0.7$. The red curve represents $b_{\rm
      tot}(k)$ for the subsample of voids whose enclosed total matter
    overdesnity was lower than the median enclosed total matter
    overdensity in the above sample. Similarly, the green curve shows
    $b_{\rm tot}(k)$ for the subsample whose enclosed total matter overdensity is higher than the median.}
  \label{split_bias}
\end{figure}

Plotting the bias defined in Eqn.\ \eqref{totalbias} in Fig. \ref{bias1}, we find that it
is indeed strongly scale dependent. To make sure that this scale
dependent linear bias is not an artifact of our simulation method, we
perform simulations of the same cosmology using two other methods. In
one, we treat the neutrinos as an additional species with a different
mass in an N-body code. In the other, we treat the neutrino fluid in a
linear approximation scheme. We then use the same criterion we have
defined above to find voids and calculate the large scale bias. We
find that the scale dependent bias we see in our method matches the
result from the N-body method to within our error bars, which denote
the scatter in the bias from different realizations. As can be seen,
treating the neutrinos in a linear approximation leads to an even
stronger scale dependence in the bias.

To illustrate how strong this effect is, we also run a set of 8 simulations with 
CDM only such that the final power spectra closely matches the final power 
spectra in the simulations with neutrinos. We then use the same void finder on 
the CDM-only runs to find voids and calculate the linear bias. This is then 
plotted on the same figure, and shows that it is extremely flat on linear 
scales.

\begin{figure}[t!]
\centering
  \includegraphics[scale=0.4]{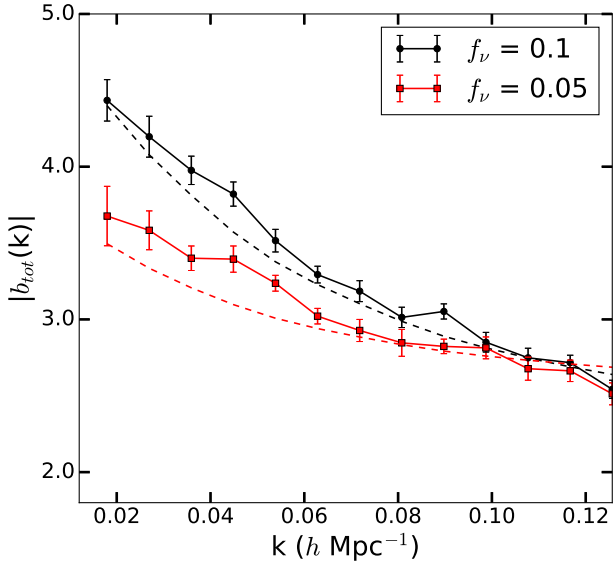}
  \caption{Effect of $f_\nu$ on the scale-dependent bias using our method for a fixed neutrino mass of $m_\nu = 0.1 \,$eV. We 
compare the case where $f_\nu = 0.1$ (black) to the case where $f_\nu=0.05$ 
(red). The dashed lines show the predictions from the spherical expansion model.}
  \label{bias2}
\end{figure}

\begin{figure}[t!]
\centering
  \includegraphics[scale=0.4]{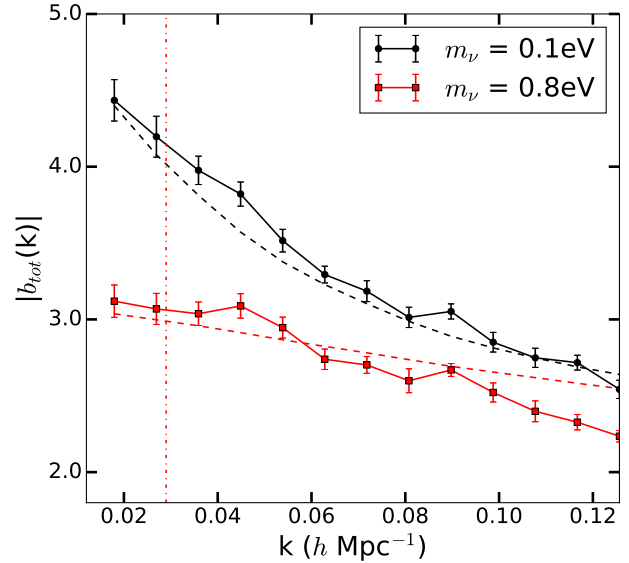}
  \caption{Effect of neutrino mass on the scale dependent bias. We compare the 
bias when the mass of the neutrino is $0.1\,$eV (black) to the case when the mass 
is $0.8\,$eV (red) with $f_\nu = 0.1$ in both cases. The red dot-dashed line gives the free streaming length for the 
$0.8$eV neutrino. The free streaming length of the lighter neutrino lies outside 
(to the left of) this plot. The scale dependent bias is evident below the 
streaming length of each species. The dashed lines show predictions from the spherical 
expansion model.}
  \label{bias3}
\end{figure}

To study the nature of the scale dependence, we perform a $\chi^2$ fit of the large 
scale bias from our simulations using Eqn.\ \eqref{scaledepbias} with
$b_{\rm CDM}(k)$ and $b_\nu(k)$ assumed to be scale independent.  For
the simulation parameters described above, we find a best-fit 
value of $b_{\rm CDM} = -1.91$ and $b_\nu = -3.69$ with $\chi^2$/d.o.f.\ value of 
$0.488$. We do a similar fit for the large scale bias obtained from simulations 
which assumed the neutrinos to be a linear fluid, which showed a stronger scale 
dependence. In this case we get best-fit values of $b_{\rm CDM} = -1.94$ and $b_\nu 
= -3.95$ with $\chi^2$/d.o.f value of $0.587$. We see $b_\nu$ shows a larger 
difference in the two cases than $b_{\rm CDM}$. Also, the absolute value of $b_\nu$ 
is larger in the linear approximation, meaning that the voids we have selected 
are rarer in the linear approximation simulations. In the linear approximation,
we truncate the evolution equations at the first order in perturbed quantities 
like the overdensities and peculiar velocities, and so creating deep voids becomes
more difficult.

To obtain an analytic understanding of these results, we use the spherical 
expansion model \citep[e.g.][]{Fillmore1984,Jennings2013}, modified to include the effects of 
massive neutrinos. Firstly, we assume that the ratio of the Lagrangian void biases 
$b^L_\nu$ and $b^L_{\rm CDM}$ is given by
\begin{equation}
 \frac{b^L_\nu}{b^L_{\rm CDM}} = \frac{f_\nu  \frac{d \Delta_{\rm nl}}{d\delta_{\rm lin}}\big |_{\nu}}
 {f_{\rm CDM}  \frac{d \Delta_{\rm nl}}{d\delta_{\rm lin}}\big |_{\rm CDM}}
 \label{bias_ratio}
 \end{equation}	
where $\Delta_{\rm nl}$ is the actual average nonlinear density of each species 
enclosed in the voids, while $\delta_{\rm lin}$ is the prediction for the average enclosed
overdensity from linear theory only. Since the neutrino perturbations are still expected to 
be small (and therefore close to expectations from linear perturbation theory) in the voids 
that we studied, we assume that $\frac{d \Delta_{\rm nl}}{d\delta_{\rm lin}}\big |_{\nu} = 1$. 

In the spherical expansion model, we assume that neutrinos form a smooth component, and
therefore expand with the background. The CDM component can be thought of as an underdense 
universe, whose evolution is governed by the Friedmann equation. The time 
evolution of the scale factor $a$ in the unperturbed background universe follows
\begin{equation}
 \left(\frac{da}{dt}\right) = H_0 \left[\frac{1}{a}\right]^{1/2}
 \label{sf}
\end{equation}
where $H_0$ is the present Hubble parameter and we have assumed $\Omega_m = 1$. The perturbed universe, whose scale factor we denote as $\chi$ evolves as 
\begin{equation}
 \left(\frac{d\chi}{d t}\right) = H_0\left[\frac{f_{\rm CDM}}{\chi} + 
 \frac{5}{3} f_{\rm CDM} |\delta| + \frac{\chi^2(1-f_{\rm CDM})}{a^3}\right]^{1/2}
 \label{pt_sf}
\end{equation}
where we have used the fact $f_\nu = 1-f_{\rm CDM}$ is a non-clustering component of the total matter density.
The nonlinear overdensity of the CDM component at a time $t$ is 
\begin{equation}
 \Delta_{\rm nl} (t) = \left(\frac{a(t)}{\chi(t)}\right)^3 - 1
 \label{delta_nl}
\end{equation}

Eqns.\ \eqref{pt_sf} and \eqref{sf} can be solved to find the average
enclosed nonlinear overdesnity $\Delta_{\rm nl}$ in Eqn.\ \eqref{delta_nl} as a function of the linear overdensity $\delta$.
This can then be substituted back into the denominator of the
right-hand side of Eqn.\ \eqref{bias_ratio}. Once we obtain the Lagrangian biases, we need a prescription to convert to
Eulerian biases to compare with the simulation results. This
mapping is not straightforward in the presence of a free streaming
species, i.e.\ the neutrinos \citep{Loverde2014}. As an approximation we assume that the
advection of the voids is dominated by the advection of the CDM
component, neglecting the effect of the advection of neutrinos. This
approximation is equivalent to setting $b_\nu = b^L_\nu$ and $b_{\rm
  CDM} = b^L_{\rm CDM} + 1$, where the value of $b_{\rm CDM}$ is calculated using the best fit to the simualation results. 
We then compare the ratio of the Eulerian biases from this calculation to the ratio we find in
simulations for different void threshold definitions. This comparison is shown in 
Fig.\ \ref{model}, which demonstrates that this simple model seems to quantitatively explain the trend seen in the simulations, for void definitions using overdensity thresholds in the
range $-0.5$ to $-0.8$. Below threshold of $-0.8$, there are too few objects in the simulations
box to get reliable statistics on the behavior of the bias. We also do not go above $-0.5$ in the 
void definition so that the voids we select always have a Lagrangian bias of $\lesssim -2$.

Since most galaxy surveys define voids using galaxy counts, which
depend on the underlying CDM field, rather than the total matter
field, we investigate the behavior of the bias when we select voids in
the simulation using the CDM field only. Once again, we used an
enclosed overdensity threshold of $-0.7$. We plot $b_{\rm tot}(k)$ for
this sample of voids with the black curve in Fig.\ \ref{split_bias} and
find that it shows very little scale dependence. From this sample of
voids selected using the CDM densities only, we split into two
subsamples based on the enclosed total matter overdensity - above and
below the median value in the sample. When we plot the behavior of
$b_{\rm tot}(k)$ for the two subsamples, (green and red curves in
Fig. \ref{split_bias}) we find a very strong scale dependence in the
biasing. This strategy of identifying voids using the CDM field and
then splitting the sample using the total matter field could be used
in surveys like DES where both galaxy counts and lensing data is
available to search for this scale dependent bias. Here we have not
taken into account the noise in the lensing signal, which will, of
course, be present in the lensing data from actual surveys. This noise
will serve to weaken the differences in the behavior of the biases of
the two samples, and in future work we will investigate the effect of
lensing noise in washing out the signal shown in Fig.\ \ref{split_bias}.

We also investigated how this scale-dependent linear bias changes as we vary 
$f_\nu$ defined in Eqn.\ \eqref{fnu}. Once again we ran 8 realizations with the 
same cosmological parameters as above, except for $\Omega_{\rm CDM}$ and 
$\Omega_\nu$, for which we used the values $0.285$ and $0.015$ respectively. 
This is equivalent to $f_\nu = 0.05$. As seen in Fig.\ref{bias2}, this leads to 
two effects - the voids defined similarly in both sets of simulations have a 
lower overall amplitude  of the bias for the $f_\nu = 0.05$ case, and secondly, the scale 
dependence also decreases.  This lower bias is a consequence of the
lower mass fraction in neutrinos: with smaller $f_\nu$, it is easier
to make more voids of the size we consider, and therefore the
magnitude of the bias of such objects decreases.

Next, we investigated how the neutrino free streaming scale affects the 
scale dependence that we observe in the void bias. To do this, we ran a set of 8 
simulations for a $m=0.8 \,$eV neutrino with $f_\nu=0.1$. We keep the other 
cosmological parameters to be the same as earlier. We use such a heavy neutrino 
species to illustrate how the shape of the void bias is different above and 
below the free streaming length of the neutrino. For the $m=0.8 \,$eV, the free 
streaming scale is about $250 h^{-1}$Mpc - this scale is represented by the 
dotted red line in Fig.\ 
\ref{bias3}. At scales larger than the free streaing scale (smaller $k$), the 
CDM power spectum and the neutrino power spectrum are the same, and there is very little  
scale dependence in the void bias. On the other hand, at scales below the free 
streaming scale where the CDM and the neutrino power spectra are significantly 
different, the bias starts showing a clear scale dependence. This is in contrast 
to the $m=0.1\,$eV neutrino, for which the free streaming scale is larger than 
our simulation box and therefore shows a scale dependent void bias at all the 
scales that we investigate. We also see that the scale dependence is steeper for 
the lighter neutrino particle - this is because the power spectrum for the 
lighter neutrino is more damped (and therefore more different from the CDM power 
spectrum) at any given scale. 

\begin{figure}[t!]
\centering
  \includegraphics[scale=0.4]{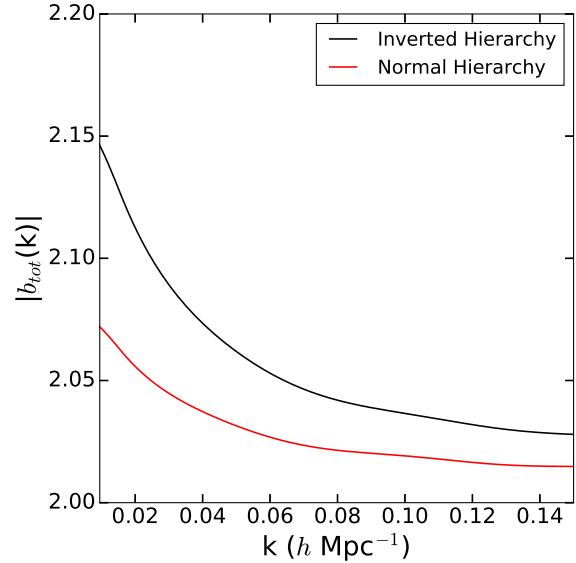}
  \caption{Scale dependent bias from the spherical expansion model for inverted hierarchy 
  (black curve) and normal hierarchy (red curve). In the inverted hierarchy,
  $\sum m_\nu = 0.12 \,$eV, $f_\nu \sim 0.0087$ and $f_{\rm CDM} \sim 0.9913$. In the inverted hierarchy,
  $\sum m_\nu = 0.06 \,$eV, $f_\nu \sim 0.0043$ and $f_{\rm CDM} \sim 0.9957$. Voids are defined as 
  regions enclosing an underdensity $-0.7$ in both cases.  We assume that 
  $b_{\rm CDM} = -2$ for reference. }
  \label{bias_ih}
\end{figure}

While we plan to do a comprehensive study of this scale dependent void bias for 
realistic ranges of neutrino masses, we can use the 
spherical expansion model calculation to predict
the approximate size of the bias effect for realistic neutrino number densities
and minimal masses in the normal hierarchy and the inverted
hierarchy. For the inverted hierarchy case, we assume the minimal sum of
the neutrino masses, $\sum m_\nu \approx 0.12 \,$eV, so that  
$\Omega_\nu \approx 0.0026$.  With the choice of $\Omega_\Lambda = 0.7$, we have
$f_{\rm CDM} \approx 0.9913$ and $f_\nu \approx 0.0087$. In the normal
hierarchy, the minimal sum of the neutrino masses is $\sum m_\nu
\approx 0.06 \,$eV, which gives $\Omega_\nu \approx 0.0013$. In this
case,  $f_{\rm CDM} \approx 0.9957$ and $f_\nu \approx 0.0043$. We use
the spherical expansion model to compute the ratio of the biases
$b_\nu$ and $b_{\rm CDM}$, defining voids to be enclosing 
a total underdensity of $-0.7$. We then used linear transfer functions and 
power spectra at $z=0$, generated using CLASS, to see how strongly
$b_{\rm tot}$ varies as a function of scale on these large scales. We
plot  the result in Fig.\ \ref{bias_ih} assuming that the voids have
$b_{\rm CDM} = -2$ for reference. From our $f_\nu = 0.1$ simulations, we found that voids with radius about $6\, h^{-1}$Mpc to $8 \,h^{-1}$Mpc with an enclosed overdensity threshold of $-0.7$ had Eulerian biases close to $-2$. 

To quickly evaluate the ratio of the biases $b^L_\nu/b^L_{\rm CDM}$ predicted by the spherical expansion 
model over the range of realistic values of the neutrino masses as a function of 
$f_\nu$ and the threshold underdensity for void definition, $\Delta_v$, we provide below a fitting 
function in these variables. We find that 
\begin{equation}
 \frac{b^L_\nu}{b^L_{\rm CDM}} \approx A f_\nu^b |\Delta|^c \exp\left(d |\Delta|\right) 
\end{equation}
with the best fit parameters $A = 5.37 \times 10^{-6}$, $b = 1.07$, $c = -7.57$ and $d = 16.77$. 
This fitting formula is accurate to a few percent over the range of $0.001 <f_\nu < 0.025$ 
and $0.5< |\Delta| < 0.8$.

\section{Conclusions}
\label{concl}

In this paper we have presented a novel, hybrid method for performing cosmological 
simulations in which particles have finite thermal velocities, using 
hydrodynamics along with standard N-body techniques used in CDM simulations. We 
have tested our method extensively in both the linear regime, as well as on 
non-linear problems.

For cosmologies with massive neutrinos, our novel method appears to
accurately evolve cosmic structures at all redshifts and all scales,
including both linear and nonlinear regimes.  This is in contrast to
traditional N-body methods, which are known to produce significant
errors in the clustering of neutrinos even at low redshift (e.g.\
Fig.\ \ref{nugrowth}).    However, for warm dark
matter (WDM) cosmologies, we find that the Eulerian version of our
method produces too much numerical diffusion to be
useful.  Instead, we find that using Lagrangian approaches 
to solving the fluid equations, like smoothed particle hydrodynamics,
allows us to circumvent the problem of diffusion on small scales.

We have also presented a novel effect in cosmologies with massive neutrinos 
- the strong scale dependence in the bias of voids, even on linear scales. These 
voids were defined using the overall matter fields (CDM and neutrinos), rather 
than just the CDM fields. We note that even for voids defined using CDM field 
only, there is scale dependence in the large scale bias, but that effect is very 
weak, similar to the dependence seen in the halo bias in 
\citet{Villaescusa2013,Loverde2014}. For mass-defined halos, the scale 
dependence of the bias found from our hybrid method matches the results from 
simulations in which neutrinos were treated as N-body particles to within the 
error bars. However, linear theory simulations over-predict the strength of the 
scale dependence.  For an inverted mass hierarchy of neutrinos, with 
$f_\nu \sim 0.01$, we find about $5\%$ scale dependence of the void
bias. For the normal hierarchy, the effect is about $2\%$.
A detailed study scanning all of the allowed parameter space in 
neutrino mass is required to predict how strongly this effect might show up in 
actual observations.
Observationally, voids are normally identified using galaxy counts in 
surveys \citep[e.g.][]{Sanchez2016,Clampitt2015}, rather than identification from
the lensing shear field.  Our results indicate that it may be
worthwhile to attempt to detect voids in the mass field directly from
lensing, rather than from galaxies, since mass-selected voids should
exhibit the neutrino-dependent bias effect discussed above. The other strategy could be to 
first identify voids using the galaxy counts, and then subdividing the sample based on the mass field, which can be inferred through the gravitational lensing signal. As we have shown, the two subsamples of voids should exhibit scale dependence in their bias. 

The simulations presented in this paper used a fixed grid and
therefore had low spatial resolution.  In the future, we plan to
combine our fluid method with adaptive mesh refinement simulation
codes, which will allow us to study dark matter halos and galaxies. A particularly
interesting area to study is redshift space distortions. The scale dependent growth 
factor in cosmologies with massive neutrinos is in principle observable in 
surveys measuring these redshift space distortions. While this effect has been 
studied using methods like perturbation theory \citep{Upadhye2015} or 
N-body techniques \citep{Marulli2011}, it will be interesting to see
how the results from our hybrid simulations compare with these other approaches.

\acknowledgements{ 
  We thank Mani Chandra, Charles Gammie, Nickolay Gnedin, Andrey
  Kravtsov, Marilena LoVerde, and Masahiro Takada for many helpful
  discussions.  We also thank Julien Lesgourgues and Thomas Tram for
  assistance with the CLASS software. This work was supported by NASA under
  grant NNX12AD02G.  ND was also supported by a Sloan Fellowship, by
  the Institute for Advanced Study, by the Ambrose Monell Foundation,
  and by the Center for Advanced Study at UIUC.  }

\newcommand{\jcap}{JCAP}
\bibliography{code}

\begin{thebibliography}{74}
\expandafter\ifx\csname natexlab\endcsname\relax\def\natexlab#1{#1}\fi

\bibitem[{Ahmad {et~al.}(2001)Ahmad, Allen, Andersen, Anglin, B\"uhler, Barton,
  Beier, Bercovitch, Bigu, Biller, Black, Blevis, Boardman, Boger, Bonvin,
  Boulay, Bowler, Bowles, Brice, Browne, Bullard, Burritt, Cameron, Cameron,
  Chan, Chen, Chen, Chen, Chon, Cleveland, Clifford, Cowan, Cowen, Cox, Dai,
  Dai, Dalnoki-Veress, Davidson, Doe, Doucas, Dragowsky, Duba, Duncan, Dunmore,
  Earle, Elliott, Evans, Ewan, Farine, Fergani, Ferraris, Ford, Fowler, Frame,
  Frank, Frati, Germani, Gil, Goldschmidt, Grant, Hahn, Hallin, Hallman, Hamer,
  Hamian, Haq, Hargrove, Harvey, Hazama, Heaton, Heeger, Heintzelman, Heise,
  Helmer, Hepburn, Heron, Hewett, Hime, Howe, Hykawy, Isaac, Jagam, Jelley,
  Jillings, Jonkmans, Karn, Keener, Kirch, Klein, Knox, Komar, Kouzes, Kutter,
  Kyba, Law, Lawson, Lay, Lee, Lesko, Leslie, Levine, Locke, Lowry, Luoma,
  Lyon, Majerus, Mak, Marino, McCauley, McDonald, McDonald, McFarlane,
  McGregor, McLatchie, Drees, Mes, Mifflin, Miller, Milton, Moffat, Moorhead,
  Nally, Neubauer, Newcomer, Ng, Noble, Norman, Novikov, O'Neill, Okada,
  Ollerhead, Omori, Orrell, Oser, Poon, Radcliffe, Roberge, Robertson,
  Robertson, Rowley, Rusu, Saettler, Schaffer, Schuelke, Schwendener, Seifert,
  Shatkay, Simpson, Sinclair, Skensved, Smith, Smith, Starinsky, Steiger,
  Stokstad, Storey, Sur, Tafirout, Tagg, Tanner, Taplin, Thorman, Thornewell,
  Trent, Tserkovnyak, Van~Berg, Van~de Water, Virtue, Waltham, Wang, Wark,
  West, Wilhelmy, Wilkerson, Wilson, Wittich, Wouters, \& Yeh}]{SNO2001}
Ahmad, Q.~R., Allen, R.~C., Andersen, T.~C., {et~al.} 2001, Phys. Rev. Lett.,
  87, 071301

\bibitem[{Ahn {et~al.}(2006)Ahn, Aliu, Andringa, Aoki, Aoyama, Argyriades,
  Asakura, Ashie, Berghaus, Berns, Bhang, Blondel, Borghi, Bouchez, Boyd,
  Burguet-Castell, Casper, Catala, Cavata, Cervera, Chen, Cho, Choi, Dore,
  Echigo, Espinal, Fechner, Fernandez, Fujii, Fujii, Fukuda, Fukuda,
  Gomez-Cadenas, Gran, Hara, Hasegawa, Hasegawa, Hayashi, Hayato, Helmer,
  Higuchi, Hill, Hiraide, Hirose, Hosaka, Ichikawa, Ieiri, Iinuma, Ikeda,
  Inagaki, Ishida, Ishihara, Ishii, Ishii, Ishino, Ishitsuka, Itow, Iwashita,
  Jang, Jang, Jeon, Jeong, Joo, Jover, Jung, Kajita, Kameda, Kaneyuki, Kang,
  Kato, Kato, Kearns, Kerr, Kim, Khabibullin, Khotjantsev, Kielczewska, Kim,
  Kim, Kim, Kim, Kim, Kitamura, Kitching, Kobayashi, Kobayashi, Kohama, Konaka,
  Koshio, Kropp, Kubota, Kudenko, Kume, Kuno, Kurimoto, Kutter, Learned,
  Likhoded, Lim, Lim, Loverre, Ludovici, Maesaka, Mallet, Mariani, Martens,
  Maruyama, Matsuno, Matveev, Mauger, McConnel~Mahn, McGrew, Mikheyev,
  Minakawa, Minamino, Mine, Mineev, Mitsuda, Mitsuka, Miura, Moriguchi, Morita,
  Moriyama, Nakadaira, Nakahata, Nakamura, Nakano, Nakata, Nakaya, Nakayama,
  Namba, Nambu, Nawang, Nishikawa, Nishino, Nishiyama, Nitta, Noda, Noumi,
  Nova, Novella, Obayashi, Okada, Okumura, Okumura, Onchi, T., Oser, Otaki,
  Oyama, Pac, Park, Pierre, Rodriguez, Saji, Sakai, Sakuda, Sakurai, Sanchez,
  Sarrat, Sasaki, Sato, Sato, Scholberg, Schroeter, Sekiguchi, Seo, Sharkey,
  Shima, Shiozawa, Shiraishi, Sitjes, Smy, So, Sobel, Sorel, Stone, Sulak,
  Suga, Suzuki, Suzuki, Suzuki, Tada, Takahashi, Takasaki, Takatsuki, Takenaga,
  Takenaka, Takeuchi, Takeuchi, Taki, Takubo, Tamura, Tanaka, Tanaka, Tanaka,
  Tanaka, Tashiro, Terri, T'Jampens, Tornero-Lopez, Toshito, Totsuka, Ueda,
  Vagins, Whitehead, Walter, Wang, Wilkes, Yamada, Yamada, Yamamoto, Yamanoi,
  Yanagisawa, Yershov, Yokoyama, Yokoyama, Yoo, Yoshida, \& Zalipska}]{K2K2003}
Ahn, M.~H., Aliu, E., Andringa, S., {et~al.} 2006, Phys. Rev. D, 74, 072003

\bibitem[{{Ali-Ha{\"i}moud} \& {Bird}(2013)}]{Yacine2013}
{Ali-Ha{\"i}moud}, Y., \& {Bird}, S. 2013, \mnras, 428, 3375, 1209.0461

\bibitem[{An {et~al.}(2012)An, Bai, Balantekin, Band, Beavis, Beriguete,
  Bishai, Blyth, Boddy, Brown, {et~al.}}]{Dayabay2012}
An, F., Bai, J., Balantekin, A., {et~al.} 2012, Physical Review Letters, 108,
  171803

\bibitem[{{Angulo} {et~al.}(2013){Angulo}, {Hahn}, \& {Abel}}]{Angulo2013}
{Angulo}, R.~E., {Hahn}, O., \& {Abel}, T. 2013, \mnras, 434, 3337, 1304.2406

\bibitem[{{Archidiacono} \& {Hannestad}(2016)}]{Archidiacono2016}
{Archidiacono}, M., \& {Hannestad}, S. 2016, \jcap, 6, 018, 1510.02907

\bibitem[{{Avila-Reese} {et~al.}(2001){Avila-Reese}, {Col{\'{\i}}n},
  {Valenzuela}, {D'Onghia}, \& {Firmani}}]{AvilaReese2001}
{Avila-Reese}, V., {Col{\'{\i}}n}, P., {Valenzuela}, O., {D'Onghia}, E., \&
  {Firmani}, C. 2001, \apj, 559, 516, astro-ph/0010525

\bibitem[{{Behroozi} {et~al.}(2012){Behroozi}, {Wechsler}, \&
  {Wu}}]{Behroozi2012}
{Behroozi}, P., {Wechsler}, R., \& {Wu}, H.-Y. 2012, {Rockstar: Phase-space
  halo finder}, Astrophysics Source Code Library

\bibitem[{Binney \& Tremaine(2011)}]{binney2011galactic}
Binney, J., \& Tremaine, S. 2011, Galactic dynamics (Princeton university
  press)

\bibitem[{{Blas} {et~al.}(2011){Blas}, {Lesgourgues}, \& {Tram}}]{Blas2011}
{Blas}, D., {Lesgourgues}, J., \& {Tram}, T. 2011, \jcap, 7, 34, 1104.2933

\bibitem[{{Bode} {et~al.}(2001){Bode}, {Ostriker}, \& {Turok}}]{Bode2001}
{Bode}, P., {Ostriker}, J.~P., \& {Turok}, N. 2001, \apj, 556, 93,
  astro-ph/0010389

\bibitem[{{Brandbyge} \& {Hannestad}(2009)}]{Brandbyge2009}
{Brandbyge}, J., \& {Hannestad}, S. 2009, \jcap, 5, 2, 0812.3149

\bibitem[{{Brandbyge} \& {Hannestad}(2010)}]{Brandbyge2010a}
---. 2010, \jcap, 1, 21, 0908.1969

\bibitem[{{Brandbyge} {et~al.}(2010){Brandbyge}, {Hannestad}, {Haugb{\o}lle},
  \& {Wong}}]{Brandbyge2010b}
{Brandbyge}, J., {Hannestad}, S., {Haugb{\o}lle}, T., \& {Wong}, Y.~Y.~Y. 2010,
  \jcap, 9, 14, 1004.4105

\bibitem[{{Carbone} {et~al.}(2016){Carbone}, {Petkova}, \&
  {Dolag}}]{Carbone2016}
{Carbone}, C., {Petkova}, M., \& {Dolag}, K. 2016, ArXiv e-prints, 1605.02024

\bibitem[{{Castorina} {et~al.}(2015){Castorina}, {Carbone}, {Bel}, {Sefusatti},
  \& {Dolag}}]{Castorina2015}
{Castorina}, E., {Carbone}, C., {Bel}, J., {Sefusatti}, E., \& {Dolag}, K.
  2015, \jcap, 7, 043, 1505.07148

\bibitem[{{Castorina} {et~al.}(2014){Castorina}, {Sefusatti}, {Sheth},
  {Villaescusa-Navarro}, \& {Viel}}]{Castorina2014}
{Castorina}, E., {Sefusatti}, E., {Sheth}, R.~K., {Villaescusa-Navarro}, F., \&
  {Viel}, M. 2014, \jcap, 2, 49, 1311.1212

\bibitem[{{Clampitt} {et~al.}(2016){Clampitt}, {Jain}, \&
  {S{\'a}nchez}}]{Clampitt2015}
{Clampitt}, J., {Jain}, B., \& {S{\'a}nchez}, C. 2016, \mnras, 456, 4425,
  1507.08031

\bibitem[{{Colberg} {et~al.}(2005){Colberg}, {Sheth}, {Diaferio}, {Gao}, \&
  {Yoshida}}]{Colberg2005}
{Colberg}, J.~M., {Sheth}, R.~K., {Diaferio}, A., {Gao}, L., \& {Yoshida}, N.
  2005, \mnras, 360, 216, astro-ph/0409162

\bibitem[{{Costanzi} {et~al.}(2013){Costanzi}, {Villaescusa-Navarro}, {Viel},
  {Xia}, {Borgani}, {Castorina}, \& {Sefusatti}}]{Costanzi2013}
{Costanzi}, M., {Villaescusa-Navarro}, F., {Viel}, M., {et~al.} 2013, \jcap,
  12, 12, 1311.1514

\bibitem[{{Dalal} {et~al.}(2008){Dalal}, {Dor{\'e}}, {Huterer}, \&
  {Shirokov}}]{Dalal2008}
{Dalal}, N., {Dor{\'e}}, O., {Huterer}, D., \& {Shirokov}, A. 2008, \prd, 77,
  123514, 0710.4560

\bibitem[{{de Blok}(2010)}]{deBlok2010}
{de Blok}, W.~J.~G. 2010, Advances in Astronomy, 2010, 5, 0910.3538

\bibitem[{{de Blok} {et~al.}(2008){de Blok}, {Walter}, {Brinks},
  {Trachternach}, {Oh}, \& {Kennicutt}}]{deBlok2008}
{de Blok}, W.~J.~G., {Walter}, F., {Brinks}, E., {et~al.} 2008, \aj, 136, 2648,
  0810.2100

\bibitem[{{Diemand} {et~al.}(2008){Diemand}, {Kuhlen}, {Madau}, {Zemp},
  {Moore}, {Potter}, \& {Stadel}}]{Diemand2008}
{Diemand}, J., {Kuhlen}, M., {Madau}, P., {et~al.} 2008, \nat, 454, 735,
  0805.1244

\bibitem[{{Diemand} \& {Moore}(2011)}]{Diemand2011}
{Diemand}, J., \& {Moore}, B. 2011, Advanced Science Letters, 4, 297, 0906.4340

\bibitem[{{Dodelson}(2003)}]{Dodelson2003book}
{Dodelson}, S. 2003, {Modern cosmology} ({Academic Press})

\bibitem[{Eguchi {et~al.}(2003)Eguchi, Enomoto, Furuno, Goldman, Hanada, Ikeda,
  Ikeda, Inoue, Ishihara, Itoh, {et~al.}}]{Kamland2003}
Eguchi, K., Enomoto, S., Furuno, K., {et~al.} 2003, Physical Review Letters,
  90, 021802

\bibitem[{{Fillmore} \& {Goldreich}(1984)}]{Fillmore1984}
{Fillmore}, J.~A., \& {Goldreich}, P. 1984, \apj, 281, 9

\bibitem[{Fukuda {et~al.}(1998)Fukuda, Hayakawa, Ichihara, Inoue, Ishihara,
  Ishino, Itow, Kajita, Kameda, Kasuga, Kobayashi, Kobayashi, Koshio, Miura,
  Nakahata, Nakayama, Okada, Okumura, Sakurai, Shiozawa, Suzuki, Takeuchi,
  Totsuka, Yamada, Earl, Habig, Kearns, Messier, Scholberg, Stone, Sulak,
  Walter, Goldhaber, Barszczxak, Casper, Gajewski, Halverson, Hsu, Kropp,
  Price, Reines, Smy, Sobel, Vagins, Ganezer, Keig, Ellsworth, Tasaka,
  Flanagan, Kibayashi, Learned, Matsuno, Stenger, Takemori, Ishii, Kanzaki,
  Kobayashi, Mine, Nakamura, Nishikawa, Oyama, Sakai, Sakuda, Sasaki, Echigo,
  Kohama, Suzuki, Haines, Blaufuss, Kim, Sanford, Svoboda, Chen, Conner,
  Goodman, Sullivan, Hill, Jung, Martens, Mauger, McGrew, Sharkey, Viren,
  Yanagisawa, Doki, Miyano, Okazawa, Saji, Takahata, Nagashima, Takita,
  Yamaguchi, Yoshida, Kim, Etoh, Fujita, Hasegawa, Hasegawa, Hatakeyama,
  Iwamoto, Koga, Maruyama, Ogawa, Shirai, Suzuki, Tsushima, Koshiba, Nemoto,
  Nishijima, Futagami, Hayato, Kanaya, Kaneyuki, Watanabe, Kielczewska, Doyle,
  George, Stachyra, Wai, Wilkes, \& Young}]{SuperK98}
Fukuda, Y., Hayakawa, T., Ichihara, E., {et~al.} 1998, Phys. Rev. Lett., 81,
  1562

\bibitem[{{Gao} {et~al.}(2005){Gao}, {Springel}, \& {White}}]{Gao2005}
{Gao}, L., {Springel}, V., \& {White}, S.~D.~M. 2005, \mnras, 363, L66,
  astro-ph/0506510

\bibitem[{Harten(1983)}]{Harten1983}
Harten, A. 1983, Journal of computational physics, 49, 357

\bibitem[{{Heitmann} {et~al.}(2005){Heitmann}, {Ricker}, {Warren}, \&
  {Habib}}]{Heitmann2005}
{Heitmann}, K., {Ricker}, P.~M., {Warren}, M.~S., \& {Habib}, S. 2005, \apjs,
  160, 28, astro-ph/0411795

\bibitem[{{Heitmann} {et~al.}(2008){Heitmann}, {Luki{\'c}}, {Fasel}, {Habib},
  {Warren}, {White}, {Ahrens}, {Ankeny}, {Armstrong}, {O'Shea}, {Ricker},
  {Springel}, {Stadel}, \& {Trac}}]{Heitmann2008}
{Heitmann}, K., {Luki{\'c}}, Z., {Fasel}, P., {et~al.} 2008, Computational
  Science and Discovery, 1, 015003, 0706.1270

\bibitem[{{Hezaveh} {et~al.}(2013){Hezaveh}, {Dalal}, {Holder}, {Kuhlen},
  {Marrone}, {Murray}, \& {Vieira}}]{Hezaveh2013}
{Hezaveh}, Y., {Dalal}, N., {Holder}, G., {et~al.} 2013, \apj, 767, 9,
  1210.4562

\bibitem[{{Hezaveh} {et~al.}(2016){Hezaveh}, {Dalal}, {Marrone}, {Mao},
  {Morningstar}, {Wen}, {Blandford}, {Carlstrom}, {Fassnacht}, {Holder},
  {Kemball}, {Marshall}, {Murray}, {Perreault Levasseur}, {Vieira}, \&
  {Wechsler}}]{Hezaveh2016}
{Hezaveh}, Y.~D., {Dalal}, N., {Marrone}, D.~P., {et~al.} 2016, \apj, 823, 37,
  1601.01388

\bibitem[{{Hobbs} {et~al.}(2016){Hobbs}, {Read}, {Agertz}, {Iannuzzi}, \&
  {Power}}]{Hobbs2016}
{Hobbs}, A., {Read}, J.~I., {Agertz}, O., {Iannuzzi}, F., \& {Power}, C. 2016,
  \mnras, 458, 468, 1503.02689

\bibitem[{Hockney \& Eastwood(1988)}]{Hockney1988}
Hockney, R.~W., \& Eastwood, J.~W. 1988, Computer simulation using particles
  (CRC Press)

\bibitem[{{Inman} {et~al.}(2015){Inman}, {Emberson}, {Pen}, {Farchi}, {Yu}, \&
  {Harnois-D{\'e}raps}}]{Inman2015}
{Inman}, D., {Emberson}, J.~D., {Pen}, U.-L., {et~al.} 2015, \prd, 92, 023502,
  1503.07480

\bibitem[{{Jain} \& {Khoury}(2010)}]{Jain2010}
{Jain}, B., \& {Khoury}, J. 2010, Annals of Physics, 325, 1479, 1004.3294

\bibitem[{{Jenkins} {et~al.}(2001){Jenkins}, {Frenk}, {White}, {Colberg},
  {Cole}, {Evrard}, {Couchman}, \& {Yoshida}}]{Jenkins2001}
{Jenkins}, A., {Frenk}, C.~S., {White}, S.~D.~M., {et~al.} 2001, \mnras, 321,
  372, astro-ph/0005260

\bibitem[{{Jennings} {et~al.}(2013){Jennings}, {Li}, \& {Hu}}]{Jennings2013}
{Jennings}, E., {Li}, Y., \& {Hu}, W. 2013, \mnras, 434, 2167, 1304.6087

\bibitem[{{Klypin} {et~al.}(1999){Klypin}, {Kravtsov}, {Valenzuela}, \&
  {Prada}}]{Klypin1999}
{Klypin}, A., {Kravtsov}, A.~V., {Valenzuela}, O., \& {Prada}, F. 1999, \apj,
  522, 82, astro-ph/9901240

\bibitem[{{Lesgourgues}(2011)}]{Lesgourgues2011a}
{Lesgourgues}, J. 2011, ArXiv e-prints, 1104.2932

\bibitem[{{Lesgourgues} \& {Pastor}(2014)}]{Lesgourgues2014}
{Lesgourgues}, J., \& {Pastor}, S. 2014, New Journal of Physics, 16, 065002,
  1404.1740

\bibitem[{{Lesgourgues} \& {Tram}(2011)}]{Lesgourgues2011b}
{Lesgourgues}, J., \& {Tram}, T. 2011, \jcap, 9, 32, 1104.2935

\bibitem[{{LoVerde}(2014)}]{Loverde2014}
{LoVerde}, M. 2014, \prd, 90, 083530, 1405.4855

\bibitem[{{LoVerde}(2016)}]{Loverde2016}
---. 2016, \prd, 93, 103526, 1602.08108

\bibitem[{{Marulli} {et~al.}(2011){Marulli}, {Carbone}, {Viel}, {Moscardini},
  \& {Cimatti}}]{Marulli2011}
{Marulli}, F., {Carbone}, C., {Viel}, M., {Moscardini}, L., \& {Cimatti}, A.
  2011, \mnras, 418, 346, 1103.0278

\bibitem[{{Massara} {et~al.}(2015){Massara}, {Villaescusa-Navarro}, {Viel}, \&
  {Sutter}}]{Massara2015}
{Massara}, E., {Villaescusa-Navarro}, F., {Viel}, M., \& {Sutter}, P.~M. 2015,
  \jcap, 11, 018, 1506.03088

\bibitem[{{Mo} {et~al.}(2010){Mo}, {van den Bosch}, \& {White}}]{Mo2010book}
{Mo}, H., {van den Bosch}, F.~C., \& {White}, S. 2010, {Galaxy Formation and
  Evolution}

\bibitem[{{Monaghan}(1992)}]{Monaghan1992}
{Monaghan}, J.~J. 1992, \araa, 30, 543

\bibitem[{{Moore} {et~al.}(1999){Moore}, {Ghigna}, {Governato}, {Lake},
  {Quinn}, {Stadel}, \& {Tozzi}}]{Moore1999}
{Moore}, B., {Ghigna}, S., {Governato}, F., {et~al.} 1999, \apjl, 524, L19,
  astro-ph/9907411

\bibitem[{{Navarro} {et~al.}(1997){Navarro}, {Frenk}, \& {White}}]{Navarro1997}
{Navarro}, J.~F., {Frenk}, C.~S., \& {White}, S.~D.~M. 1997, \apj, 490, 493,
  astro-ph/9611107

\bibitem[{{Peebles}(1980)}]{Peebles1980}
{Peebles}, P.~J.~E. 1980, {The large-scale structure of the universe}

\bibitem[{{Planck Collaboration} {et~al.}(2014){Planck Collaboration}, {Ade},
  {Aghanim}, {Armitage-Caplan}, {Arnaud}, {Ashdown}, {Atrio-Barandela},
  {Aumont}, {Baccigalupi}, {Banday}, \& et~al.}]{Planck2013}
{Planck Collaboration}, {Ade}, P.~A.~R., {Aghanim}, N., {et~al.} 2014, \aap,
  571, A16, 1303.5076

\bibitem[{Roe(1986)}]{Roe1986}
Roe, P. 1986, Annual review of fluid mechanics, 18, 337

\bibitem[{{S{\'a}nchez} {et~al.}(2016){S{\'a}nchez}, {Clampitt}, {Kovacs},
  {Jain}, {Garc{\'{\i}}a-Bellido}, {Nadathur}, {Gruen}, {Hamaus}, {Huterer},
  {Vielzeuf}, {Amara}, {Bonnett}, {DeRose}, {Hartley}, {Jarvis}, {Lahav},
  {Miquel}, {Rozo}, {Rykoff}, {Sheldon}, {Wechsler}, {Zuntz}, {Abbott},
  {Abdalla}, {Annis}, {Benoit-L{\'e}vy}, {Bernstein}, {Bernstein}, {Bertin},
  {Brooks}, {Buckley-Geer}, {Carnero Rosell}, {Carrasco Kind}, {Carretero},
  {Crocce}, {Cunha}, {D'Andrea}, {da Costa}, {Desai}, {Diehl}, {Dietrich},
  {Doel}, {Evrard}, {Fausti Neto}, {Flaugher}, {Fosalba}, {Frieman},
  {Gaztanaga}, {Gruendl}, {Gutierrez}, {Honscheid}, {James}, {Krause}, {Kuehn},
  {Lima}, {Maia}, {Marshall}, {Melchior}, {Plazas}, {Reil}, {Romer}, {Sanchez},
  {Schubnell}, {Sevilla-Noarbe}, {Smith}, {Soares-Santos}, {Sobreira},
  {Suchyta}, {Tarle}, {Thomas}, {Walker}, \& {Weller}}]{Sanchez2016}
{S{\'a}nchez}, C., {Clampitt}, J., {Kovacs}, A., {et~al.} 2016, ArXiv e-prints,
  1605.03982

\bibitem[{{Scherrer} \& {Weinberg}(1998)}]{Scherrer1998}
{Scherrer}, R.~J., \& {Weinberg}, D.~H. 1998, \apj, 504, 607, astro-ph/9712192

\bibitem[{{Springel}(2005)}]{Springel2005}
{Springel}, V. 2005, \mnras, 364, 1105, astro-ph/0505010

\bibitem[{{Springel} {et~al.}(2005){Springel}, {White}, {Jenkins}, {Frenk},
  {Yoshida}, {Gao}, {Navarro}, {Thacker}, {Croton}, {Helly}, {Peacock}, {Cole},
  {Thomas}, {Couchman}, {Evrard}, {Colberg}, \& {Pearce}}]{Springel2005b}
{Springel}, V., {White}, S.~D.~M., {Jenkins}, A., {et~al.} 2005, \nat, 435,
  629, astro-ph/0504097

\bibitem[{{Springel} {et~al.}(2008){Springel}, {Wang}, {Vogelsberger},
  {Ludlow}, {Jenkins}, {Helmi}, {Navarro}, {Frenk}, \& {White}}]{Springel2008}
{Springel}, V., {Wang}, J., {Vogelsberger}, M., {et~al.} 2008, \mnras, 391,
  1685, 0809.0898

\bibitem[{{Stadel} {et~al.}(2009){Stadel}, {Potter}, {Moore}, {Diemand},
  {Madau}, {Zemp}, {Kuhlen}, \& {Quilis}}]{Stadel2009}
{Stadel}, J., {Potter}, D., {Moore}, B., {et~al.} 2009, \mnras, 398, L21,
  0808.2981

\bibitem[{{Tinker} {et~al.}(2008){Tinker}, {Kravtsov}, {Klypin}, {Abazajian},
  {Warren}, {Yepes}, {Gottl{\"o}ber}, \& {Holz}}]{Tinker2008}
{Tinker}, J., {Kravtsov}, A.~V., {Klypin}, A., {et~al.} 2008, \apj, 688, 709,
  0803.2706

\bibitem[{{Tinker} {et~al.}(2010){Tinker}, {Robertson}, {Kravtsov}, {Klypin},
  {Warren}, {Yepes}, \& {Gottl{\"o}ber}}]{Tinker2010}
{Tinker}, J.~L., {Robertson}, B.~E., {Kravtsov}, A.~V., {et~al.} 2010, \apj,
  724, 878, 1001.3162

\bibitem[{Toro(2009)}]{Toro2009}
Toro, E.~F. 2009, Riemann solvers and numerical methods for fluid dynamics: a
  practical introduction (Springer Science \& Business Media)

\bibitem[{Upadhye {et~al.}(2016)Upadhye, Kwan, Pope, Heitmann, Habib, Finkel,
  \& Frontiere}]{Upadhye2015}
Upadhye, A., Kwan, J., Pope, A., {et~al.} 2016, Phys. Rev. D, 93, 063515

\bibitem[{Van~Leer(1977)}]{Vanleer1977}
Van~Leer, B. 1977, Journal of computational physics, 23, 276

\bibitem[{{Villaescusa-Navarro} {et~al.}(2013){Villaescusa-Navarro}, {Bird},
  {Pe{\~n}a-Garay}, \& {Viel}}]{Villaescusa2013}
{Villaescusa-Navarro}, F., {Bird}, S., {Pe{\~n}a-Garay}, C., \& {Viel}, M.
  2013, \jcap, 3, 19, 1212.4855

\bibitem[{{Villaescusa-Navarro} \& {Dalal}(2011)}]{Villaescusa2011b}
{Villaescusa-Navarro}, F., \& {Dalal}, N. 2011, \jcap, 3, 24, 1010.3008

\bibitem[{{Villaescusa-Navarro} {et~al.}(2014){Villaescusa-Navarro}, {Marulli},
  {Viel}, {Branchini}, {Castorina}, {Sefusatti}, \& {Saito}}]{Villaescusa2014}
{Villaescusa-Navarro}, F., {Marulli}, F., {Viel}, M., {et~al.} 2014, \jcap, 3,
  11, 1311.0866

\bibitem[{{Villaescusa-Navarro} {et~al.}(2011){Villaescusa-Navarro},
  {Vogelsberger}, {Viel}, \& {Loeb}}]{Villaescusa2011}
{Villaescusa-Navarro}, F., {Vogelsberger}, M., {Viel}, M., \& {Loeb}, A. 2011,
  ArXiv e-prints, 1106.2543

\bibitem[{{Wang} \& {White}(2007)}]{Wang2007}
{Wang}, J., \& {White}, S.~D.~M. 2007, \mnras, 380, 93, astro-ph/0702575

\bibitem[{{Warren} {et~al.}(2006){Warren}, {Abazajian}, {Holz}, \&
  {Teodoro}}]{Warren2006}
{Warren}, M.~S., {Abazajian}, K., {Holz}, D.~E., \& {Teodoro}, L. 2006, \apj,
  646, 881, astro-ph/0506395

\bibitem[{{Zemp}(2009)}]{Zemp2009}
{Zemp}, M. 2009, Modern Physics Letters A, 24, 2291, 0909.4298

\end{thebibliography}
\bibliographystyle{apj}
\end{document}